\documentclass[twocolumn]{aastex63}
\usepackage{amsmath,amsthm,amsfonts,amssymb,bm}
\usepackage{physics,commath}

\usepackage{color}
\usepackage{multirow}
\usepackage{listings}
\hypersetup{breaklinks}
\usepackage{graphicx,comment}

\usepackage{textcomp}
\usepackage{epstopdf}
\usepackage{natbib}
\usepackage{algorithm,algcompatible}
\definecolor{royalblue}{rgb}{0.25, 0.41, 0.88}
\definecolor{brickred}{rgb}{0.8, 0.25, 0.33}
\hypersetup{colorlinks=True,linkcolor={royalblue},citecolor={royalblue},urlcolor={brickred},breaklinks=true}

\usepackage{lineno}

\makeatletter
\newcommand*\bigcdot{\mathpalette\bigcdot@{.5}}
\newcommand*\bigcdot@[2]{\mathbin{\vcenter{\hbox{\scalebox{#2}{$\m@th#1\bullet$}}}}}

\newcommand{\argmin}{\mathop{\rm arg~min}\limits}

\DeclareMathOperator{\arccosh}{arccosh}

\renewcommand\labelenumi{(\roman{enumi})}
\renewcommand\theenumi\labelenumi

\newcommand{\xlrv}[1]{#1}

\algnewcommand\INPUT{\item[\textbf{Input:}]}
\algnewcommand\OUTPUT{\item[\textbf{Output:}]}

\makeatother

\begin{document}
\title{Three-Dimensional Reconstruction of Weak-Lensing Mass Maps \\
with a Sparsity Prior. I. Cluster Detection}
\author{Xiangchong Li}
\email{xiangchong.li@ipmu.jp}
\affiliation{Department of Physics, University of Tokyo, Tokyo 113-0033, Japan}
\affiliation{Kavli Institute for the Physics and Mathematics of the Universe
(WPI), University of Tokyo, Chiba 277-8583, Japan}
\author{Naoki Yoshida}
\affiliation{Department of Physics, University of Tokyo, Tokyo 113-0033, Japan}
\affiliation{Kavli Institute for the Physics and Mathematics of the Universe
(WPI), University of Tokyo, Chiba 277-8583, Japan}
\affiliation{Institute for Physics of Intelligence, University of Tokyo, Tokyo
113-0033, Japan}
\author{Masamune Oguri}
\affiliation{Department of Physics, University of Tokyo, Tokyo 113-0033, Japan}
\affiliation{Kavli Institute for the Physics and Mathematics of the Universe
(WPI), University of Tokyo, Chiba 277-8583, Japan}
\affiliation{Research Center for the Early Universe, University of Tokyo, Tokyo
113-0033, Japan}
\author{Shiro Ikeda}
\affiliation{The Institute of Statistical Mathematics, Tokyo 190-8562, Japan}
\affiliation{Department of Statistical Science, Graduate University for
Advanced Studies, Tokyo 190-8562, Japan}
\author{Wentao Luo}
\affiliation{Kavli Institute for the Physics and Mathematics of the Universe
(WPI), University of Tokyo, Chiba 277-8583, Japan}
\affiliation{CAS Key Laboratory for Research in Galaxies and Cosmology,
University of Science and Technology of China, Hefei, Anhui 230026, China}

\begin{abstract}
We propose a novel method to reconstruct high-resolution three-dimensional mass
maps using data from photometric weak-lensing surveys. We apply an adaptive
LASSO algorithm to perform a sparsity-based reconstruction on the assumption
that the underlying cosmic density field is represented by a sum of
Navarro--Frenk--White halos. We generate realistic mock galaxy shear catalogs
by considering the shear distortions from isolated halos for the configurations
matched to the Subaru Hyper Suprime-Cam Survey with its photometric redshift
estimates. We show that the adaptive method significantly reduces line of sight
smearing that is caused by the correlation between the lensing kernels at
different redshifts. Lensing clusters with lower mass limits of $10^{14.0}
\text{h}^{-1}M_{\odot}$, $10^{14.7} \text{h}^{-1}M_{\odot}$, $10^{15.0}
\text{h}^{-1}M_{\odot}$ can be detected with 1.5-$\sigma$ confidence at the low
($z<0.3$), median ($0.3\leq z< 0.6$) and high ($0.6\leq z< 0.85$) redshifts,
respectively, with an average false detection rate of 0.022 deg$^{-2}$. The
estimated redshifts of the detected clusters are systematically lower than the
true values by $\Delta z \sim 0.03$ for halos at $z\leq 0.4$, but the relative
redshift bias is below $0.5\%$ for clusters at $0.4<z\leq 0.85$. The standard
deviation of the redshift estimation is $0.092$. Our method enables direct
three-dimensional cluster detection with accurate redshift estimates.
\end{abstract}


\section{Introduction}

Weak gravitational lensing causes small, coherent distortion of the shapes of
distant galaxies. Information on the foreground mass distribution is imprinted
in the distorted galaxy images, and thus weak-lensing offers a direct  physical
probe into the mass distribution in our universe, including both visible matter
and invisible dark matter \citep[see][for recent
reviews]{revKilbinger15,revRachel17}. Ongoing and future observational programs
such as the Subaru Hyper Suprime-Cam survey \citep[HSC;][]{HSC1-data}, the
Kilo-Degree Survey \citep[KiDS;][]{KIDS13}, the Dark Energy Survey
\citep[DES;][]{DES05}, Vera C. Rubin Observatory's Legacy Survey of Space and
Time \citep[LSST;][]{LSSTOverviwe2019}, Euclid \citep{Euclid2011}, and the
Nancy Grace Roman Space Telescope \citep{WFIRST15} are aimed at studying
large-scale mass distribution with high precision.

Statistics of density peaks in two-dimensional ($2$D) and three-dimensional
($3$D) mass maps can be used as a powerful cosmological probe
\citep{WL-massMap-peakcounts-Jain2000,WL-massMap-peakcountsAna-Fan2010,WL-massMap-peakcountsFM-Lin2016}.
One can detect massive clusters of galaxies by identifying high signal-to-noise
(S/N) peaks in a mass map without any reference to the mass-to-light ratio
\citep{WL-massMap-clusDet-Schneider1996,WL-massMap-clusDet-Hamana2004}.

$2$D density  reconstruction techniques recover integration of the projected
mass along the line of sight, \xlrv{and} have been extensively studied so far
\citep{massMap-KS1993,WL-massMap-Glimpse2D-Lanusse2016,sparseBaysianMassMap-Price2020}
and applied to large-scale surveys
\citep{HSC1-massMaps,massMapDES-Chang2018,DES-SV-massMap-sparsity}. Cluster
detection and identification from $2$D mass maps has been applied to wide-field
weak-lensing surveys
\citep{WL-massMap-clusDet-CFHT-Shan2012,WL-massMap-clusDet-HSC-Miyazaki2018,WL-massMap-clusDet-HSC-Hamana2020}.
Cross-matching with another cluster catalogs (e.g., an optically selected one)
is usually needed to infer physical quantities such as mass and to estimate
redshift for individual clusters located in $2$D mass maps.

In principle, one can directly reconstruct $3$D mass distributions by using
photometric redshift (photo-$z$) information of the source galaxies
\citep{massMap3D-Hu2002,massMap3D-Bacon2003,HST-massMap-Massey2007,LSS-massMap-Wiener-Simon2009,WL-massMap-VanderPlas2011}.
Unfortunately, these methods either do not have enough spatial resolution to
identify individual clusters, or suffer from smearing along the line of sight.
These are critical obstacles that need to be overcome for practical searches of
clusters in $3$D mass maps. Alternatively, \citet{WL-clusDet-Hennawi2005}
propose to perform a maximum-likelihood detection of clusters, by convolving
tomographic shear measurements with $3$D filters that match the tangential
shears induced by multiscale Navarro--Frenk--White (NFW) halos. Their method
can be used effectively to  detect clusters, but does not fully reconstruct
wide-field mass distributions.

In the present paper, we develop a novel method for high-resolution $3$D
reconstruction. We model a given $3$D density field as a sum of the NFW
\citep{halo-NFW1997ApJ} basis `atoms' that are setup in {\it comoving}
coordinates. A basis atom is defined by a $2$D NFW surface density profile on
the transverse plane and one-dimensional Dirac delta function in the
line of sight direction. We apply the adaptive LASSO algorithm
\citep{AdaLASSO-Zou2006} to find a sparse solution for a pixelized map. We
examine the performance of cluster detection using the reconstructed mass maps.
To this end, we apply shear distortions generated by isolated halos using
realistic HSC-like galaxy shapes with photo-$z$ estimates.

The rest of the paper is organized as follows. In Section~\ref{sec_Method}, we
propose the new method for $3$-D density map reconstruction. In
Section~\ref{sec_Test}, we study the cluster detection from the reconstructed
mass map using isolated halo simulations with the HSC observational condition.
In Section~\ref{sec_Sum}, we summarize and discuss the future development of
the method.

Throughout the present paper, we adopt the $\Lambda$CDM cosmology of the final
full-mission Planck observation of the cosmic microwave background with
$H_0=67.4 ~\rm{km~s^{-1} Mpc^{-1}}$, $\Omega_m=0.315$, $\Omega_\Lambda=0.685$,
and $\sigma_8=0.811$, $n_s=0.965$ \citep{cmb-Planck2018-Cosmology}.

\section{Method}
\label{sec_Method}
\newcommand{\MLZ}{\texttt{MLZ}}

The lensing shear field $\gamma$ observed from background galaxy images is
related to the foreground density contrast field $\delta= \rho /\bar{\rho}-1$
through a linear transform:
\begin{equation}\label{eq-delta2gamma}
    \gamma=\mathbf{T} \cdot \delta + \epsilon,
\end{equation}
where $\epsilon$ is the shear measurement error due to the random orientation
of galaxy shapes (shape noise) and the sky variance (photon noise).
\xlrv{
The  matrix operator $\mathbf{T}= \mathbf{P} \cdot \mathbf{Q}$ includes the
physical lensing effect denoted by a matrix operator $\mathbf{Q}$ and the
systematic effects in observations represented by a matrix operator
$\mathbf{P}$. The latter includes smoothing of the shear field in the
transverse plane and photometric redshift uncertainties.
}

In this section, we first introduce the NFW dictionary to model the density
contrast field (Section~\ref{subsec_method_dictionary}). Then we explain the
weak-lensing operator $\mathbf{Q}$ in Section~\ref{subsec_method_delta2shear},
whereas the systematics operator $\mathbf{P}$ is introduced in
Section~\ref{subsec_method_Systematics}. Reconstruction of the density contrast
field is performed by solving a linear problem of Equation
(\ref{eq-delta2gamma}) or of the equivalent Equation (\ref{eq-x2gamma}) in its
practical form introduced later in this section. To this end, we devise an
adaptive LASSO algorithm that achieves high-resolution reconstruction in
Section~\ref{subsec_method_reconstruction}.

\subsection{Model dictionary}
\label{subsec_method_dictionary}

\begin{figure*}[!t]
\begin{center}
\includegraphics[width=1.\textwidth]{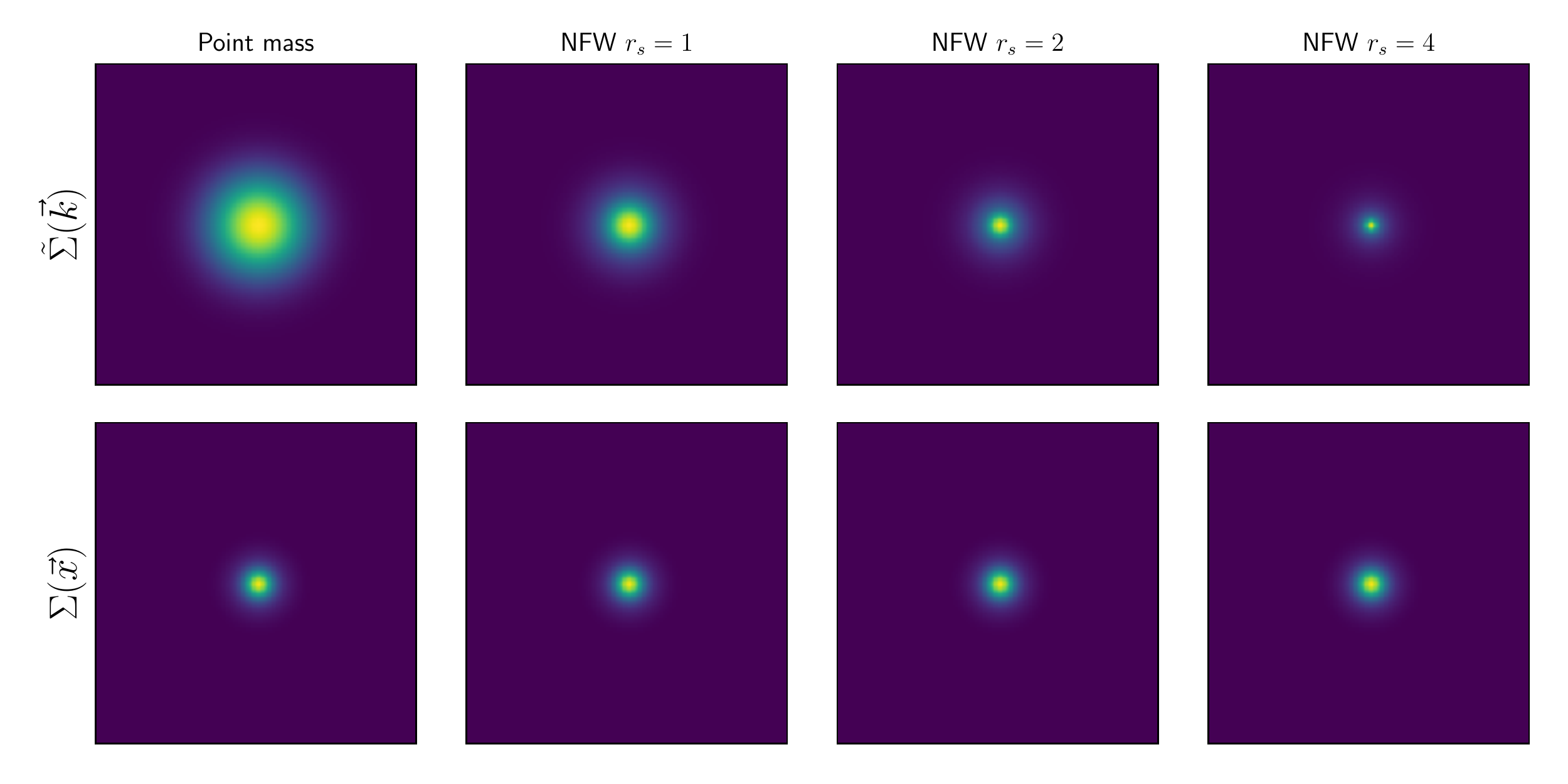}
\end{center}
\caption{
    The normalized $2$D profiles of the smoothed basis ``atoms''. The pixel
    size is $1\arcmin$. The leftmost column is the point-mass atom, and the
    other columns show the NFW atoms with different scale radii (in pixels) as
    indicated. The upper (lower) panels show the basis atoms in Fourier
    (configuration) space. We smooth the $2$D profiles using a Gaussian kernel
    with a $1.5$ pixel width.
    }
    \label{fig_atoms2D}
\end{figure*}

\begin{figure}
\begin{center}
    \includegraphics[width=0.45\textwidth]{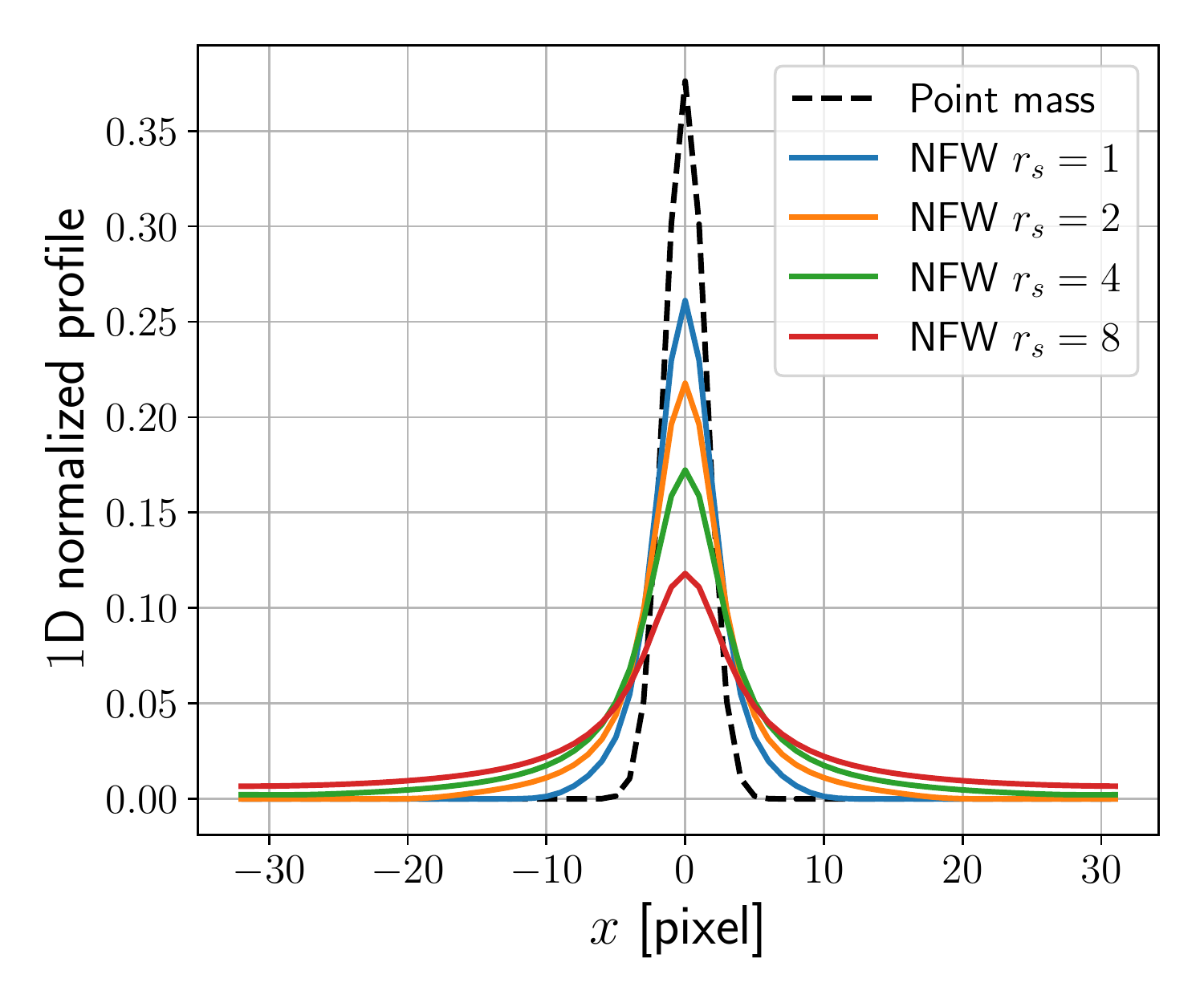}
\end{center}
\caption{
    The normalized one-dimensional ($1$D) profiles of smoothed basis atoms,
    which are slices of the corresponding $2$D profiles (shown in Figure
    \ref{fig_atoms2D}) at $y=0$. The scale radii are in pixels.
    }
    \label{fig_atoms1D}
\end{figure}

In order to reconstruct high-resolution, high-S/N mass maps, we incorporate
prior information on the density contrast field into the reconstruction by
modeling the density field as a sum of basis atoms in a ``dictionary'':
\begin{equation}\label{eq-x2delta}
    \delta= \mathbf{\Phi} \cdot x,
\end{equation}
where $\mathbf{\Phi}$ is the matrix operator transforming from the projection
coefficient vector $x$ to the density contrast field $\delta$. The column
vectors of $\mathbf{\Phi}$ are the basis ``atoms'' of the model dictionary and
denoted as $\phi_s$.

There have been a few studies that adopt different dictionaries for $3$D
weak-lensing map reconstructions:
(i) \citet{LSS-massMap-Wiener-Simon2009} perform a reconstruction in Fourier
space, which is equivalent to representing the density contrast field with
sinusoidal functions. However, sinusoidal functions are not localized in
configuration space, and the density contrast field is not sparse in Fourier
space. Therefore, sparsity priors cannot be directly applied to this model
dictionary for high-resolution mass map reconstructions.
(ii) \citet{LSS-massMap-Glimpse3D-Leonard2014} model the density contrast field
with starlets \citep{Starlet-Starck2015}. However, the starlet dictionary does
not account for the angular scale difference at different lens redshifts, and
is not specifically designed to model the clumpy mass distribution in the
universe. It is worth exploring other dictionaries for our purpose of
weak-lensing mass reconstruction.

In the standard cosmological model, dark matter is concentrated in roughly
spherical ``halos'', which have the NFW density profile
\citep{halo-NFW1997ApJ}. Motivated by this fact, we generate a model dictionary
using NFW atoms with $N$ typical scale radii $r_s~ (s=1,...,N)$ in comoving
coordinates. An atom has the NFW surface density profile on the transverse
plane \citep{haloModel-TJ2003-3pt} and the Dirac $\delta$ function in the
line of sight direction. Following \citet{LSS-massMap-Glimpse3D-Leonard2014},
we neglect the size of halos along the line of sight since the resolution scale
of the reconstruction is much larger than the halos.

The multiscale NFW atom is defined as
\begin{equation}
\phi_s(\vec{r}_{\theta},z) = \frac{f}{2 \pi r_s^2 }\,
    F(|\vec{r}_{\theta}|/r_s) \,\delta_D(z),
\end{equation}
where $\vec{r}_\theta$ is the projection of the comoving position on the
transverse plane,
\begin{equation}
F(x)=
\begin{cases}
    -\frac{\sqrt{c^2-x^2}}{(1-x^2)(1+c)} + \frac{\arccosh
    \left(\frac{x^2+c}{x(1+c)}\right)}{(1-x^2)^{3/2}}  & (x<1),\\
    \frac{\sqrt{c^2-1}}{3(1+c)} (1+\frac{1}{c+1}) & (x=1),\\
    -\frac{\sqrt{c^2-x^2}}{(1-x^2)(1+c)} +
        \frac{\arccos\left(\frac{x^2+c}{x(1+c)}\right)}
        {(x^2-1)^{3/2}} & (1<x\leq c),\\
    0& (x>c)\,.
\end{cases}
\end{equation}
$f=1/[\ln (1+c)-c/(1+c)]$, and $c$ is the halo concentration. For simplicity,
we fix $c=4$ for the NFW atoms in our dictionary. In the present paper, we
adopt a hard truncation on the NFW profile at a radius equals $cr_s$.
Studying the influence of different truncation forms
\citep{haloTrunc-Oguri2011} on the mass map reconstruction is left to our
future work.

Compared to other model dictionaries, our NFW dictionary is motivated by a
physical consideration on the clumpy mass distribution in the universe.
Furthermore, the multiscale NFW atoms are set up in comoving coordinates, with
an account of the scale difference in the angular coordinates for halos at
different lens redshifts. The corresponding NFW atom in the angular separation
coordinates is
\begin{equation}
\begin{split}
\phi_s(\vec{\theta},z) = \frac{f \chi^2 (z)}{2 \pi r_s^2 } \,
    F(|\vec{\theta}|\chi(z)/r_s) \,\delta_D(z),
\end{split}
\end{equation}
where $\chi(z)$ is the comoving distance to redshift $z$.

For the NFW dictionary, the transform from the projection
coefficient vector to the
density contrast field of Equation~\eqref{eq-x2delta} is written as
\begin{equation}\label{eq-x2delta-NFW}
\delta(\vec{r}) = \sum_{s=1}^{N} \int d^3 r'
    \phi_s(\vec{r}-\vec{r'}) x_s(\vec{r'})\,.
\end{equation}
To simplify the notation, we compress the projection coefficient vectors on the
basis atoms with multiple scales into a single column vector:
\begin{equation}
    x=\begin{pmatrix}
    x_{0}\\
    x_{1}\\
    ...\\
    x_{N}
    \end{pmatrix},
\end{equation}
and write the dictionary transform operator into a row vector:
\begin{equation}
\mathbf{\Phi}=\begin{pmatrix}
    \int {\rm d}^3r\,\phi_0(\vec{r}) ~\int {\rm d}^3r\,
    \phi_1(\vec{r})~ ...~\int {\rm d}^3r\, \phi_{N}(\vec{r})
\end{pmatrix}.
\end{equation}

For an additional test and comparison, we also construct a dictionary with the
point-mass atoms which are represented by the $3$D Dirac function as
\begin{equation}
    \phi_{\rm{PM}}(\vec{\theta},z) =
    \delta_{\rm D}(\theta_1) \,\,\delta_{\rm D}(\theta_2) \,\,\delta_{\rm D}(z)\,.
\end{equation}
The $2$D profiles of the NFW atoms and the point-mass atom on the transverse
plane are shown in Figure~\ref{fig_atoms2D}. The corresponding $1$D profiles
are plotted in Figure~\ref{fig_atoms1D}. The profiles are smoothed with a
Gaussian kernel and pixelized on linearly spaced grids. Details on the
smoothing and pixelizing operations are described in
Section~\ref{subsec_method_smoothing}.

We define the forward transform operator $\mathbf{A}= \mathbf{P} \cdot
\mathbf{Q} \cdot \mathbf{\Phi}$ where $\mathbf{P}$ represents systematic
effects and $\mathbf{Q}$ represents the physical lensing effect. With
Equations~\eqref{eq-delta2gamma} and \eqref{eq-x2delta}, we write the transform
from the coefficient vector $x$ to the observed lensing shear field as
\begin{equation}\label{eq-x2gamma}
    \gamma=\mathbf{A} \cdot x + \epsilon\,.
\end{equation}

\subsection{Weak gravitational lensing}
\label{subsec_method_delta2shear}

\begin{figure}[!t]
\begin{center}
\includegraphics[width=.45\textwidth]{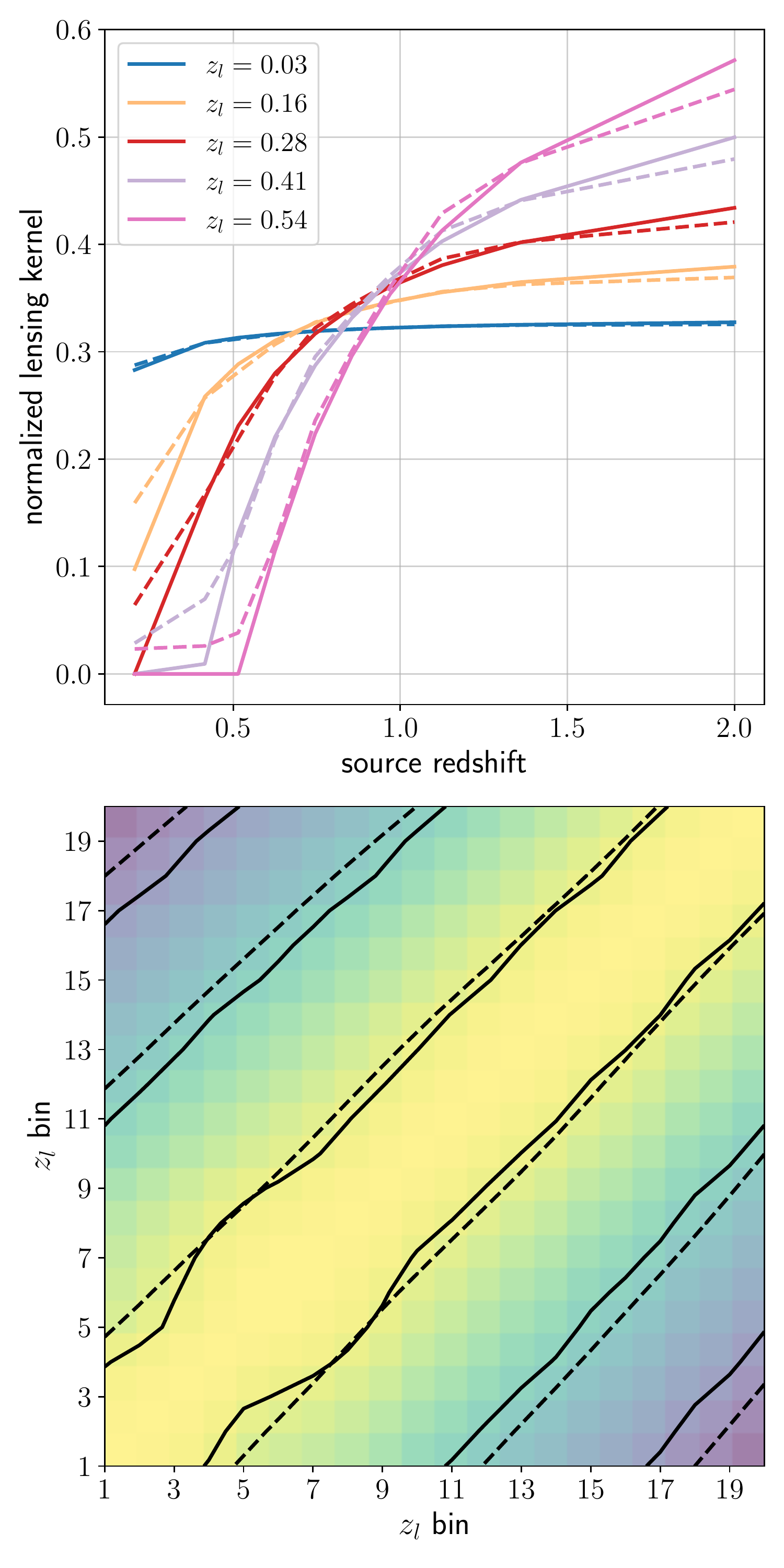}
\end{center}
\caption{
    {\em Top panel:}
    Normalized lensing kernels as a function of source redshift with lens
    redshifts fixed. The solid lines are the kernels for the source galaxies
    with precise spectroscopic redshifts, whereas the dashed lines are for
    source redshifts with HSC-like photometric redshift errors.
    {\em Bottom panel:}
    Correlation matrix between the lensing kernels of different lens redshifts.
    We normalize the diagonal terms to $1\,$. The color map is the correlation
    matrix for spectroscopic redshift. We also compare the result for the
    lensing kernel of spectroscopic (photometric) redshift by solid (dashed)
    contours at levels $0.7$, $0.85$, and $0.98\,$.
    }
    \label{fig_corlensKer}
\end{figure}

The lensing transform operator can be expressed as
\begin{equation}\label{eq-lensKerOperator}
\mathbf{Q}=\int_0^{z_s} dz_l\,  K(z_l,z_s) \int d^2 \theta'\,
    D(\vec{\theta}-\vec{\theta'})\,.
\end{equation}
$K(z_l,z_s)$ is the lensing kernel between lens redshift $z_l$ and source
redshift $z_s$ \citep{WL-rev-Bartelmann01}, which is given by
\begin{equation}
K(z_l,z_s) =
\begin{cases}
    \frac{3H_0\Omega_{\rm m}}{2 c} \frac{\chi_l \chi_{sl}
    (1+z_l)}{\chi_{s} E\left(z_l\right)} & (z_s>z_l),\\
0&(z_s \leq z_l),
\end{cases}
\end{equation}
where $E(z)$ is the Hubble parameter as a function of redshift, in units of
$H_0$.
\begin{equation}
    D(\vec{\theta})= -\frac{1}{\pi}(\theta_1-i\theta_2)^{-2}
\end{equation}
is the Kaiser-Squares kernel \citep{massMap-KS1993}, which decays proportional
to the inverse-square of the distance. Here $\theta_{1,2}$ are the two
components of the angular position vector $\vec{\theta}$.

\xlrv{
The top panel of Figure~\ref{fig_corlensKer} shows the lensing kernels as a
function of source redshift for lenses at different lens redshifts. The bottom
panel of Figure~\ref{fig_corlensKer} shows the correlation between the lensing
kernels. Each lensing kernel is highly nonlocal, and the lensing kernels of
different lens redshifts are strongly correlated as can be seen in the bottom
panel. These inherent properties of the lensing kernels result in strong
correlation between the column vectors that constitute the forward transform
matrix $\mathbf{A}$. As will be discussed in
Section~\ref{subsec_method_reconstruction}, the strong correlation makes it
challenging to reconstruct mass maps with high-resolution in the line of sight
direction.
}

\subsection{Systematics}
\label{subsec_method_Systematics}

\xlrv{
Shear measurement deviates from the true, physical shear owing to a variety of
systematic effects in real observations. In this section, we discuss the
influence of several major systematics on the lensing shear measurement, and
describe the corresponding transform operator by decomposing into three parts
of $\mathbf{R}$ (photometric redshift uncertainties), $\mathbf{W}$ (smoothing),
and $\mathbf{M}$ (masking).
}

\subsubsection{Photometric redshift Uncertainty}
\label{subsec_method_photoz}

\begin{figure}
\centering
\includegraphics[width=0.45\textwidth]{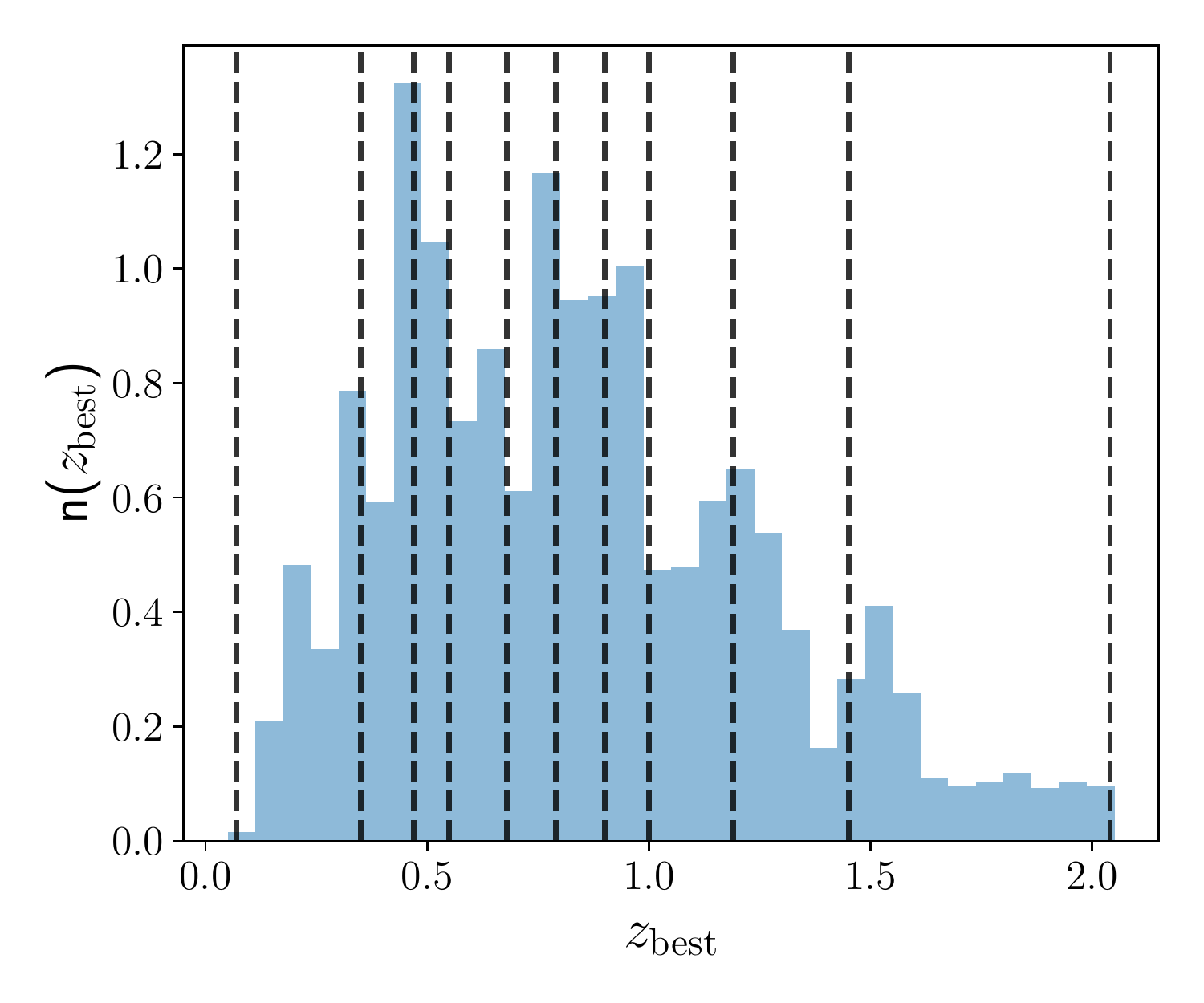}
\caption{
    The blue histogram shows the normalized number distribution of the best-fit
    photo-$z$ ($\MLZ$) estimates from tract $9347$ of the first-year HSC data.
    We use an equal-number binning scheme to divide the source galaxies into a
    total of $10$ redshift bins that are indicated by the vertical dashed
    lines.
    }
    \label{fig_bestpz}
\end{figure}

\begin{figure}
\centering
\includegraphics[width=0.45\textwidth]{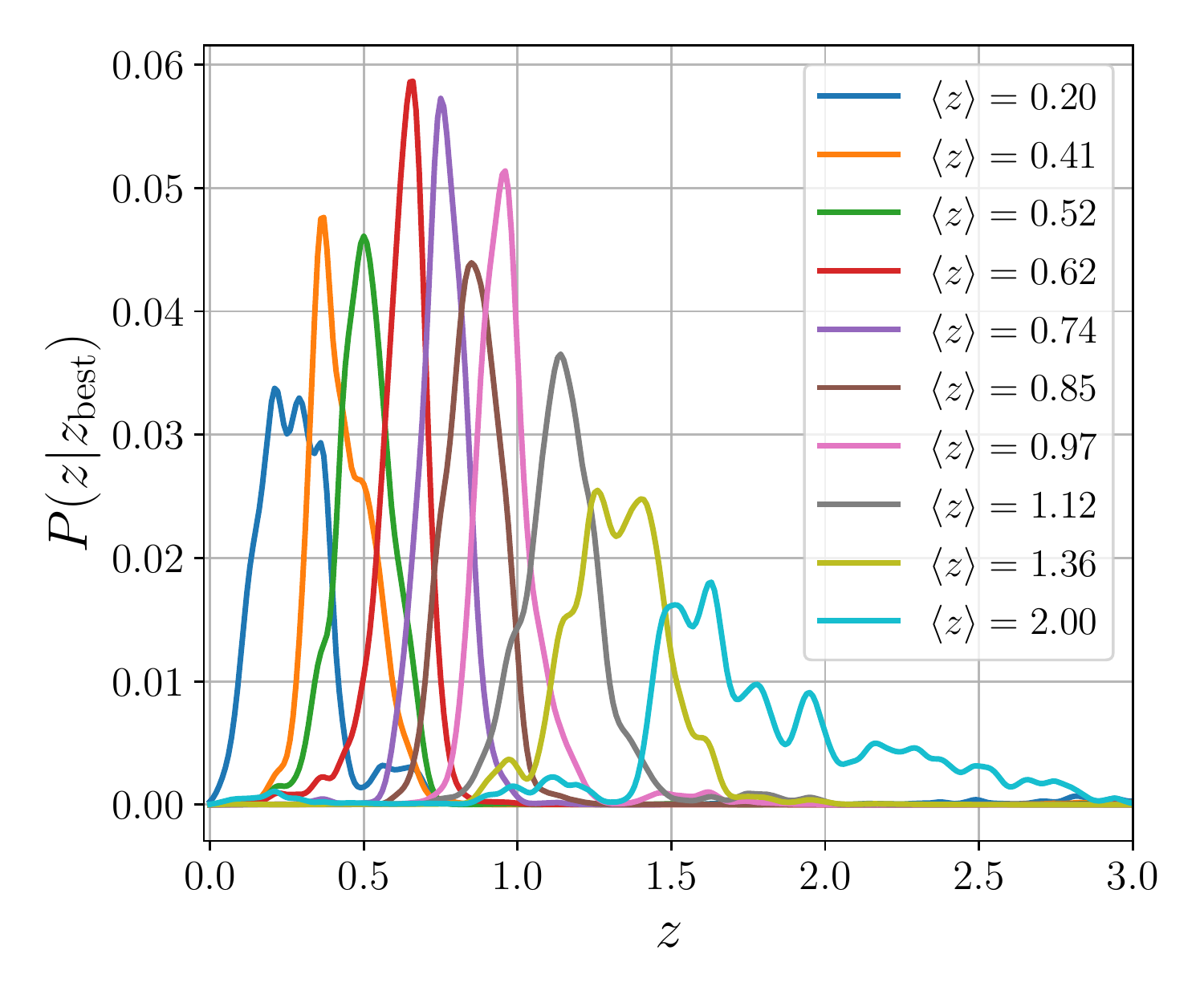}
\caption{
    The average PDF of $\MLZ$ in $10$ source redshift bins with boundaries
    defined by the vertical dashed lines in Figure~\ref{fig_bestpz}.
    }
    \label{fig_pdfpz}
\end{figure}

Photometric redshifts (photo-$z$) are estimated with a limited number of broad
optical and near-infrared bands in the current generation weak-lensing surveys.
For example, $9$ bands are used for KIDS$+$VIKING survey
\citep{KIDS_VIKING-Hildebrant2020}, $5$ bands for DES survey and HSC survey.
Unlike the high-precision spectroscopic redshift (spect-$z$) estimation, the
photo-$z$ estimation suffers from considerable statistical uncertainties.

Figure~\ref{fig_bestpz} shows the histogram of the best-fit estimates of the
$\texttt{M}$achine $\texttt{L}$earning for photo-$\texttt{Z}$
\citep[$\MLZ$;][]{MLZ-TPZ2013}
algorithm\footnote{https://github.com/mgckind/MLZ} for galaxies in
tract\footnote{The HSC data is divided into rectangular regions called tracts
covering approximately $1.7 \times 1.7 \deg^2$, and each tract is broken into
$9$x$9$ patches.} $9347$ from the photo-$z$ catalog \citep{HSC1-photoz} of the
first-year HSC data release \citep{HSC1-data}. The galaxies are divided into
ten bins according to the photo-$z$ best-fit estimates. Figure~\ref{fig_pdfpz}
shows the average probability density function (PDF) for galaxies in each
redshift bin.

\xlrv{
In the presence of photo-$z$ uncertainties, a source galaxy with a best-fit
photo-$z$ estimate $z_s$ has a posterior probability $P(z|z_s)$ of being
actually located at redshift $z$. This means that there is a possibility
$P(z|z_s)$ that the galaxy image is actually distorted by the shear at redshift
$z$. Therefore, the photo-$z$ uncertainty smears the lensing kernel {\it
statistically}, which we model by a smearing operator
}
\begin{equation}
\mathbf{R} = \int dz P(z|z_s)\,.
\end{equation}

Figure~\ref{fig_corlensKer} shows the lensing kernels and their correlations
for source redshifts with photo-$z$ uncertainties demonstrated in
Figure~\ref{fig_pdfpz}. Compared to the lensing kernels of spect-$z$ that
converge to zero at source redshifts lower than the lens redshift, the lensing
kernels of photo-$z$ do not converge to zero at the low source redshifts. This
is simply because the galaxies with photo-$z$ estimated lower than the lens
redshifts may actually be located at higher redshifts. In addition, the
photo-$z$ uncertainty only slightly increases the correlations between lensing
kernels at different lens plane.

\subsubsection{Smoothing}
\label{subsec_method_smoothing}

\begin{figure}
\begin{center}
    \includegraphics[width=0.45\textwidth]{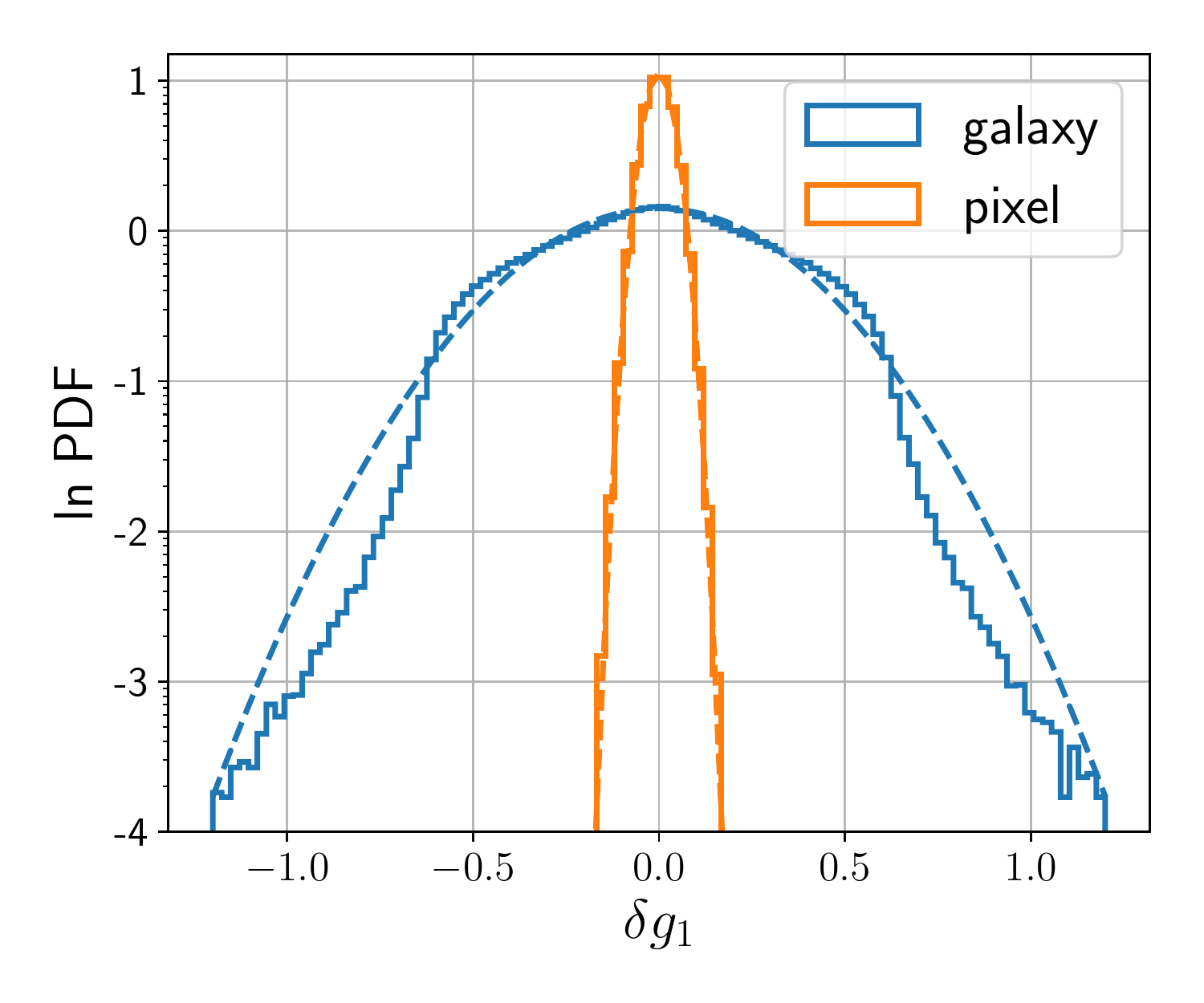}
\end{center}
\caption{
    The solid blue (orange) line shows the histogram of the HSC-like shear
    measurement error on the first component of shear $g_1$ on individual
    galaxy (pixel) level. The dashed lines are the best-fit Gaussian
    distributions to the corresponding histograms.
    }
    \label{fig_noiseHistogram}
\end{figure}

The source galaxies are not uniformly distributed in the sky, with substantial
fluctuations in the number density. We first smooth the lensing shear
measurements, and pixelize the smoothed shear field onto a regular grid. After
these procedures, the fast Fourier transform (FFT) can be directly conducted on
the transverse plane in each redshift bin to compute the convolution operation
in Equation~\eqref{eq-delta2gamma}. Another benefit of smoothing is that it
reduces bias arising from the aliasing effect in the pixelization process since
the smoothing kernel reduces the amplitude of the shear signal at high
frequency.

We convolve the lensing shear measured from galaxy images with a $3$D smoothing
kernel $W(\vec{\theta},z)$. The expectation of the lensing shear field after
smoothing is
\begin{equation}
\gamma_{\rm{sm}} (\vec{\theta},z)  = \frac{\sum_i
    W(\vec{\theta}-\vec{{\theta}}_i,z-z_i) \gamma_L(\vec{\theta}_i,z_i)}{\sum_i
    W(\vec{\theta}-\vec{{\theta}}_i,z-z_i)},
\end{equation}
where $z_i$ and $\theta_i$ refer to the photometric redshift, and transverse
position of the $i$th galaxy in the galaxy sample, respectively. $\gamma_L$
refers to the shear field before the smoothing.

The smoothing kernel $W(\vec{\theta},z)$ can be decomposed into a transverse
component $W_t(\vec{\theta})$ and a line of sight component $W_l (z)$ as
\begin{equation}
W(\vec{\theta},z)=W_t(\vec{\theta}) \,W_{l} (z)\,.
\end{equation}
We use an isotropic $2$D Gaussian kernel and a $1$D top-hat kernel to smooth
the shear field in the transverse plane and in the line of sight direction,
respectively. These components are given by
\begin{equation}
\begin{split}
W_t(\vec{\theta}) &=\frac{1}{2\pi\beta^2}\exp(-\frac{|\vec{\theta}|}{2\beta^2}),\\
W_l (z) &=
    \begin{cases}
    1/\Delta z& (|z|<\Delta z/2),\\
    0& {\rm othewise},
    \end{cases}
\end{split}
\end{equation}
where we set $\beta = 1.5$ arcmin in this paper. By definition, the smoothing
kernel is normalized as
\begin{equation}
\int {\rm d}^3 r \,W(\vec{r})=1\,.
\end{equation}
Assuming that the galaxy number distribution varies slowly at the smoothing scale, the
smoothed shear can be written into
\begin{equation}
    \gamma_{\rm{sm}} = \mathbf{W} \cdot \gamma_L,
\end{equation}
where $\mathbf{W}$ is the smoothing operator defined as
\begin{equation}
\mathbf{W} = \int {\rm d}^3 r' \, W(\vec{r}-\vec{r'})\,.
\end{equation}

We note that another widely used scheme is to average the shear measurements in
each pixel. Such a scenario is equivalent to resampling the shear field
smoothed with a $3$D top-hat kernel with the same scale as the pixels.

After smoothing, we pixelize the shear field on $N_x \times N_y \times N_z$
grids, where $N_x$ and $N_y$ are the number of pixels of the two orthogonal
axes of the transverse plane, and $N_z$ is the number of pixels in the
line of sight direction. We denote by $\gamma_{\alpha}$ the smoothed shear
measurements recorded on the pixel with index $\alpha$, where $1 \leq \alpha
\leq N_x \times N_y \times N_z$. The grids on the transverse planes are
equally spaced with a pixel size of $1\arcmin$. In the line of sight direction,
we set binning with equal galaxy number as shown in Figure~\ref{fig_bestpz}.

Similarly, we pixelize the projection coefficient vector $x$ onto an $N_x
\times N_y \times N_l$ grid. In this paper, the projection coefficient vector
is pixelized in equal spacing ranging from redshift $0.01$ to redshift $0.85$.
Here, we use $N_l$ to denote the number of the lens planes and $x_{\beta}$ to
denote the projection field element with index $1 \leq \beta \leq N_x \times
N_y \times N_l \times N$, where the last $N$ is the number of NFW dictionary
frames representing different physical scale radii. The corresponding pixelized
elements of the forward transform matrix $\mathbf{A}$ is denoted as
$A_{\alpha\beta}$.

\xlrv{
In order to see the effect of smoothing clearly, we plot the histograms of the
statistical shear measurement errors due to shape noise and sky variance for
the galaxies in tract $9347$ of the first-year HSC shear catalog
\citep{HSC1-catalog}. In Figure~\ref{fig_noiseHistogram}, we compare the error
on an individual galaxy basis and that of individual pixels after the smoothing
and pixelization procedures described in this section. The standard deviation
of the individual pixel errors is much smaller than for the galaxies owing to
the smoothing. We also note that, even though the shear measurement error for
the galaxies does not follow a Gaussian distribution, the error after smoothing
and pixelization is well described by a Gaussian distribution, which is simply
explained by the central limit theorem.
}

\subsubsection{Masking}
\label{subsec_method_msknoise}

In real observations, shear measurements can be performed in a finite region of
the sky, and the boundaries often have complicated geometries. Moreover, many
isolated sub-regions near the bright stars are masked out.

We define the masking window function as
\begin{equation}
M(\vec{r})=
\begin{cases}
1 & n_{\rm{sm}}\geq 1,\\
0 & \rm{otherwise},
\end{cases}
\end{equation}
where $n_{\rm{sm}}$ is the smoothed galaxy number density. We define the
masking operator as
\begin{equation}
\mathbf{M}= \int {\rm d}^3 r' \,M(\vec{r'}) \delta_D(\vec{r}-\vec{r'}),
\end{equation}
where $\delta_D(\vec{r})$ is $3$D Dirac delta function.

Taking into account the systematics discussed in previous
Sections~\ref{subsec_method_photoz}--\ref{subsec_method_msknoise}, the
systematic operator is
\begin{equation}\label{eq-operatorP-def}
\mathbf{P}  =  \mathbf{M} \cdot \mathbf{W} \cdot \mathbf{R}\,.
\end{equation}

\subsection{Density map reconstruction}
\label{subsec_method_reconstruction}

\begin{figure*}[!t]
\begin{minipage}[c]{1.0\columnwidth}
\begin{center}
    \includegraphics[width=1.0\textwidth]{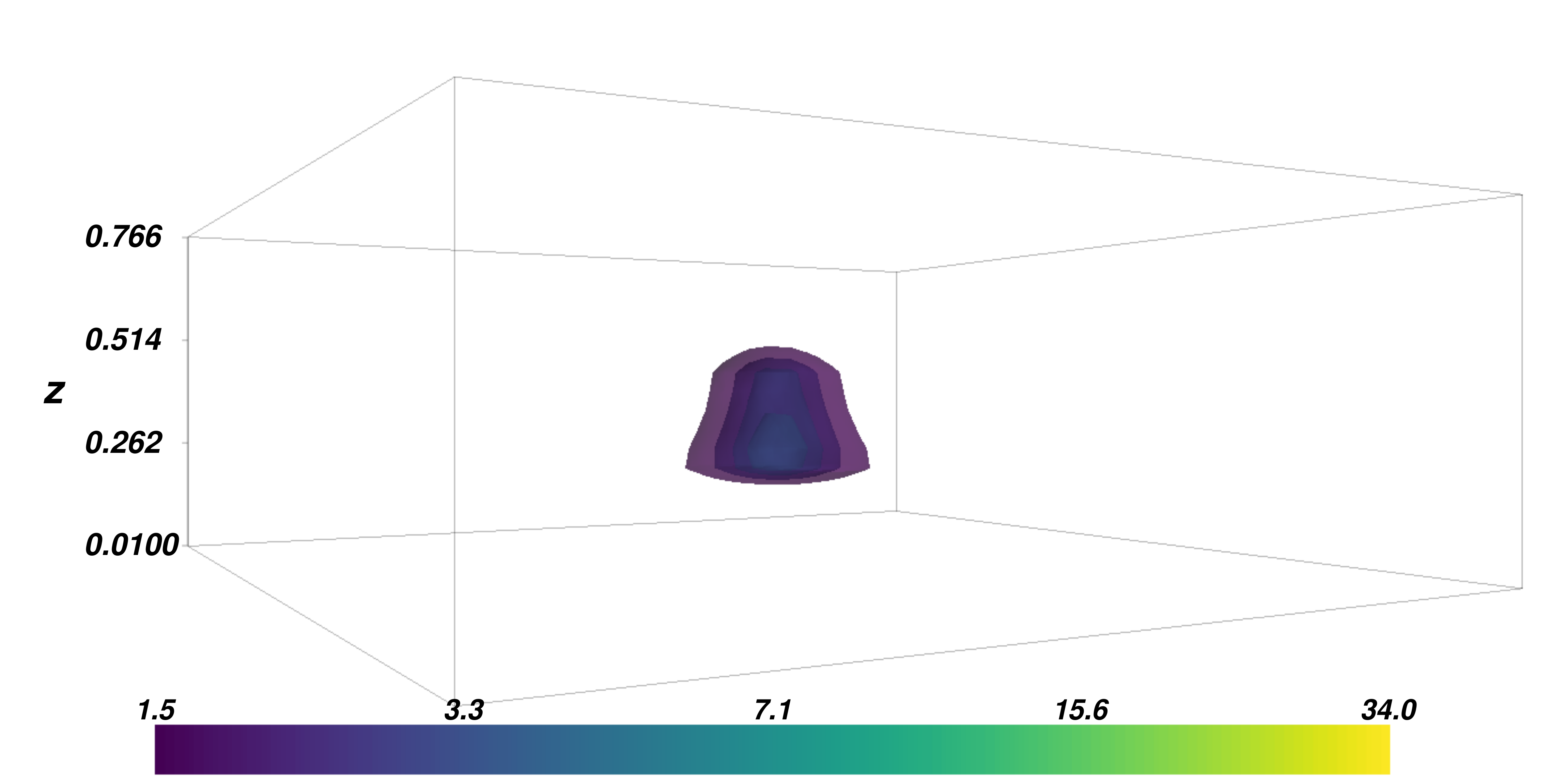}
\end{center}
\end{minipage}
\begin{minipage}[c]{1.0\columnwidth}
\begin{center}
    \includegraphics[width=1.0\textwidth]{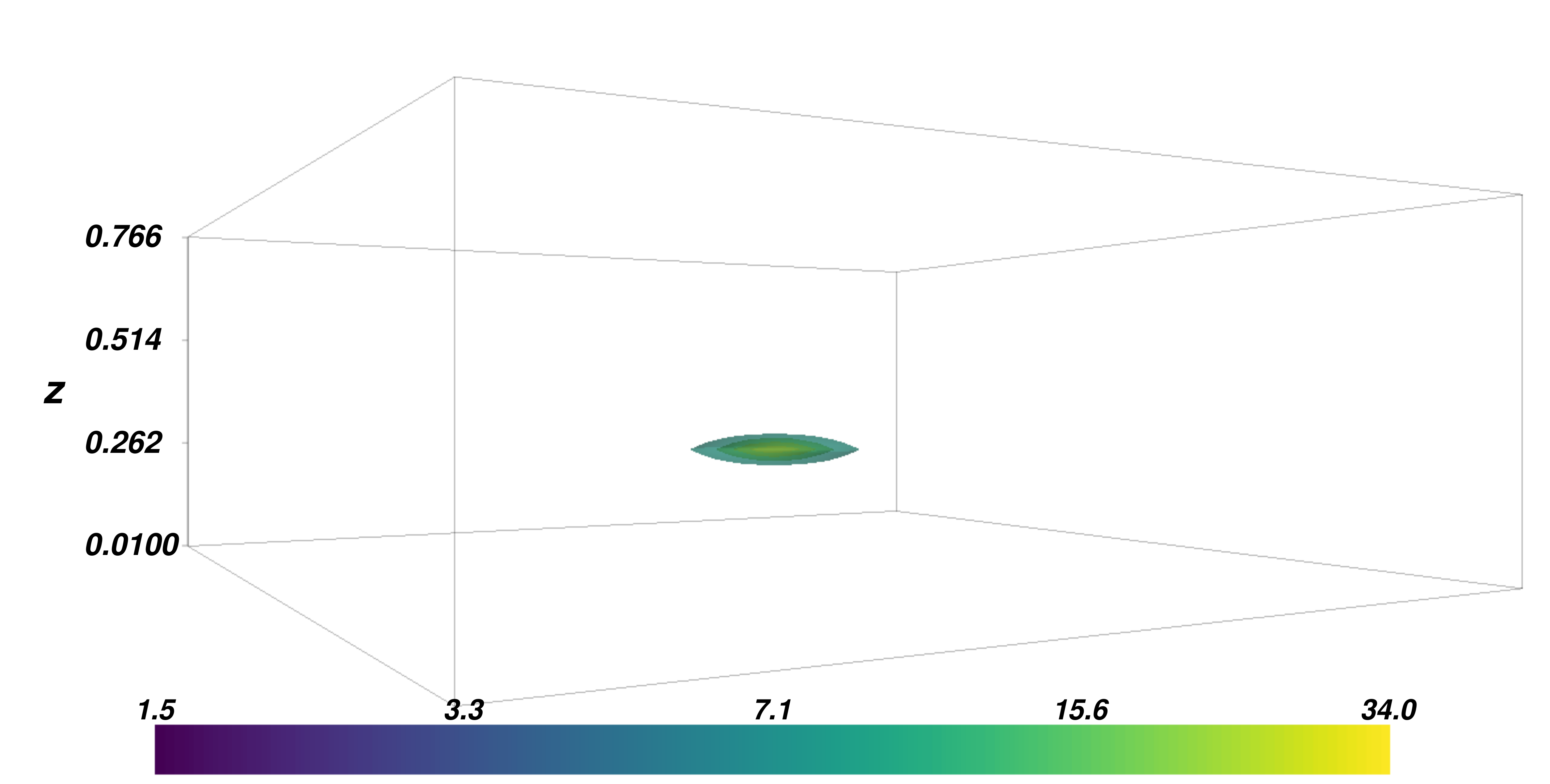}
\end{center}
\end{minipage}
\caption{
    The density map reconstruction with LASSO (left) and with our adaptive
    LASSO (right) algorithm. The $z$-axes are for redshift.  The mass of halo
    is $M_{200}=10^{15} ~\text{h}^{-1}M_{\odot}$, and its redshift is $z=0.35$. The
    color bars indicate the values of reconstructed density contrast.  Note
    that our adaptive LASSO mitigates elongation in the $z$-direction (along
    line of sight) and suppresses the shrinkage in LASSO, and thus the
    reconstructed density values are large, compared to the LASSO result shown
    in the left. Shear measurement errors and photo-$z$ uncertainties are not
    considered in this test simulation.
    }
    \label{fig_LASSOVsadaLasso}
\end{figure*}

The projection coefficients can be estimated by optimizing a penalized loss
function. An estimator is generally defined as
\begin{equation}\label{eq-lossFun_pre}
    \hat{x}=\argmin_{x} \left\{ \frac{1}{2} {}_\Sigma\norm{(\gamma-
\mathbf{A} x)}_2^2+ \lambda C(x) \right\},
\end{equation}
where ${}_\Sigma\norm{(\gamma - \mathbf{A} x)}_2^2$ is the $l^2$ norm of
residuals weighted by the inverse of the covariance matrix $\Sigma$ of the
shear measurement error, which measures the difference between the prediction
and the data. $C(x)$ is the penalty term measuring the deviation of the
coefficient estimate $x$ from the prior assumption. The estimation with the
``penalty'' term prefers parameters that are able to describe the observation
with a specified prior information. The penalty parameter $\lambda$ adjusts the
relative weight between the data and the prior assumption in the optimization
process.

There have been a few studies that adopt different penalties for $3$D
weak-lensing map reconstructions:
(i) \citet{LSS-massMap-Wiener-Simon2009} propose to use the Wiener filter,
which is also known as $l^2$ ridge penalty ($C=\norm{x}^2_2$), to find a
penalized solution in Fourier space. \citet{HSC1-massMaps} apply the method of
\citet{LSS-massMap-Wiener-Simon2009} to the first-year data of the Hyper
Suprime-Cam Survey \citep{HSC1-data}. It is found that the density maps
reconstructed by the method suffer from significant line of sight smearing with
standard deviation of $\sigma_z = 0.2 - 0.3$.
(ii) \citet{LSS-massMap-Glimpse3D-Leonard2014} use GLIMPSE algorithm, which
adopts a derivative version of the $l^1$ LASSO penalty ($C=\norm{x}^1_1$) to
find a sparse solution in the starlet dictionary \citep{Starlet-Starck2015}.
GLIMPSE reduces the smearing by adopting a ``greedy'' coordinate descent
algorithm that forces the structure to grow only on the most related lens
redshift plane.

In this paper, we use another derivative version of the LASSO penalty -- the
adaptive LASSO penalty. We first normalize the column vectors of the forward
transform matrix in Section~\ref{subsec_method_norm}. We then introduce the
loss function with the adaptive LASSO penalty in
Section~\ref{subsec_method_adaLasso}. Finally, we find the minimum of the loss
function in Section~\ref{subsec_method_FISTA} with the FISTA algorithm
\citep{FISTA-Beck2009}.

\subsection{Normalization}
\label{subsec_method_norm}

The $l^2$ norm of the $i$th column vector of $\mathbf{A}$ weighted by the
inverse of the noise covariance matrix is defined as
\begin{equation}
    \mathcal{N}_{i}= (\Sigma^{-1})_{\alpha\beta}
    A_{i\alpha}A_{i\beta}\,.
\end{equation}
The column vectors have different weighted $l^2$ norms. Since the gradient
descent algorithm, which will be used to solve Equation~\eqref{eq-lossFun_pre}
in Section~\ref{subsec_method_FISTA}, takes each column vector equally, we
normalize the column vectors before performing the density map reconstruction
to boost the convergence speed of the gradient descent iteration. The
normalized forward transform matrix and projection parameters are given by
\begin{equation}
\begin{split}
    A'_{\alpha\beta}&=A_{\alpha\beta}/
        \mathcal{N}_{\alpha}^{\frac{1}{2}},\\
    x'_{\beta}&=x_{\beta}\mathcal{N}_{\beta}^{\frac{1}{2}}\,.
\end{split}
\end{equation}

\subsubsection{Adaptive LASSO}
\label{subsec_method_adaLasso}

The LASSO algorithm uses $l^1$ norm of the projection coefficient vector to
regularize the modeling. The estimator is defined as
\begin{equation}
\hat{x'}^{\rm{LASSO}}=\argmin_{x} \left\{
\frac{1}{2} {}_\Sigma\norm{(\gamma- \mathbf{A'} \cdot x')}_2^2+
    \lambda \norm{x'}^1_1\right\},
\end{equation}
where $\norm{\bigcdot}_1^1$ and $\norm{\bigcdot}_2^2$ refer to the $l^1$ norm
and $l^2$ norm, respectively, and $\lambda$ is the penalty parameter for the
LASSO estimation.

The LASSO algorithm searches and selects the parameters that are relevant to
the measurements, and simultaneously estimates the values of the selected
parameters. It has been shown by \citet{AdaLASSO-Zou2006} that when the column
vectors of the forward transform matrix $\mathbf{A'}$ are highly correlated,
the algorithm cannot select the relevant atoms from the dictionary
consistently. In addition, the estimated parameters are often biased owing to
the shrinkage in the LASSO regression. We note that, for the density map
reconstruction problem here, the column vectors are highly correlated even in
the absence of photo-$z$ uncertainties since the lensing kernels for lenses at
different redshifts overlap significantly, i.e. highly correlated as shown in
Figure~\ref{fig_corlensKer}. Therefore, the LASSO algorithm cannot precisely
determine the consistent mass distribution in redshift, and the reconstructed
map suffers from smearing in the line of sight direction even in the absence of
noises.

Figure~\ref{fig_LASSOVsadaLasso} shows an example reconstruction results for a
single halo with mass $M_{200}=10^{15} ~\text{h}^{-1}M_{\odot}$ at redshift
$0.35$\footnote{We set the critical over-density to $200$, and use $M_{200}$ to
denote the halo mass.}. The shear measurement and photo-$z$ uncertainties are
not included in this simulation. We find significant smearing of the mass
distribution with the LASSO algorithm (left panel of
Figure~\ref{fig_LASSOVsadaLasso}).

To overcome the problem, \citet{AdaLASSO-Zou2006} proposes the adaptive LASSO
algorithm, which uses adaptive weights to penalize different projection
coefficients in the $l^1$ penalty. The adaptive LASSO algorithm performs a
two-step process. In the first step, the standard (nonadaptive) LASSO is used
to estimate the parameters. Let us denote the preliminary estimation as
$\hat{x'}_{\rm{LASSO}}$. In the second step, the preliminary estimate is used
to calculate the nonnegative weight vector for penalization as
\begin{equation}
\hat{w}= \frac{1}{\abs{\hat{x'}_{\rm{LASSO}}}^\tau},
\end{equation}
where we set the hyperparameter $\tau$ to $2\,$. The adaptive LASSO estimator
is then given by
\begin{equation}\label{eq-lossFun}
\hat{x'}=\argmin_{x'} \left\{ \frac{1}{2} {}_\Sigma\norm{(\gamma-\mathbf{A'} \cdot x')}^2_2 +
    \lambda_{\rm{ada}} \norm{\hat{w} \circ x'}^1_1\right\},
\end{equation}
where ``$\circ$'' refers to the element-wise product. $\lambda_{\rm{ada}}$ is
the penalty parameter for the adaptive LASSO, which does not need to be the
same as the penalty parameter for the preliminary LASSO estimation $\lambda$.
The adaptive weights enhance the shrinkage in the soft thresholding for the
coefficients with smaller amplitudes, whereas the weights suppress the
shrinkage for the coefficients with larger amplitudes.

To simplify the equations in the following, we rewrite the loss function with
the Einstein notation as
\begin{equation}
\begin{split}
L(x')&=\frac{1}{2}(\Sigma^{-1})_{\alpha\beta}(\gamma^{*}_{\alpha}-A'^{*}_{\alpha i}x'_{i})
    (\gamma_{\beta}-A'_{\beta j}x'_{j})\\
&+ \lambda_{\rm{ada}} \hat{w_\beta} \abs{x'_\beta},
\end{split}
\end{equation}
and the quadruple term in the loss function as
\begin{equation}
G(x')=\frac{1}{2}\Sigma^{-1}_{\alpha\beta}
    (\gamma^{*}_{\alpha}-A'^{*}_{\alpha i}x'_{i}) (\gamma_{\beta}-A'_{\beta j}x'_{j})\,.
\end{equation}

\subsubsection{FISTA}
\label{subsec_method_FISTA}

\citet{FISTA-Beck2009} propose the Fast Iterative Soft Thresholding Algorithm
(FISTA) to solve the LASSO problem. Since the loss functions of the LASSO and
the adaptive LASSO differ only in their penalty terms, FISTA is also applicable
to solve the adaptive LASSO problem in a straightforward manner. In this paper,
we apply FISTA to solve both the preliminary LASSO estimation and the adaptive
LASSO estimation.

We first explain the LASSO preliminary estimation. The coefficients are
initialized as $x_i^{(1)}=0$. According to FISTA, we iteratively update the
projection coefficient vector $x$. Taking the $n$th iteration as an example, a
temporary update is first calculated as
\begin{equation}
    x'^{(n+1)}_{i}=\mathrm{ST}_{\lambda}
    \left(x'^{(n)}_{i} -\mu \partial_i G(x'^{(n)})\right),
\end{equation}
where $\mathrm{ST}$ is the soft thresholding function defined as
\begin{equation}
\mathrm{ST}_{\lambda} \left(x'\right) =
    \mathrm{sign} (x') \max \left\{\abs{x'}-\lambda,0\right\}\,.
\end{equation}
The soft thresholding is a part of the LASSO algorithm, which selects the modes
with amplitude greater than $\lambda$, and shrinks the selected estimation by
$\lambda$.
The coefficient $\mu$ is the step size of the gradient descent iteration, and
$\partial_i G(x'^{(n)})$ is the $i$th element of the gradient vector of $G$ at
point $x'^{(n)}$:
\begin{equation}
\partial_i G(x'^{(n)})=\Sigma^{-1}_{\alpha\beta}
    \Re\left\{A'^{*}_{\alpha i}(\gamma_{\beta}-A'_{\beta j}x'_{j})\right\},
\end{equation}
where $\Re\left\{\bullet \right\}$ returns the real part of the input vector.
FISTA requires an additional update using a weighted average between
$x'^{(n+1)}$ and $x'^{(n)}$ as
\begin{equation}
\begin{split}
t^{(n+1)}&=\frac{1+\sqrt{1+4(t^{(n)})^2}}{2},\\
x'^{(n+1)} &\leftarrow x'^{(x+1)}+ \frac{t^{(n)}-1}{t^{(n+1)}}(x'^{(n+1)}-x'^{(n)}),
\end{split}
\end{equation}
where the relative weight is initialized as $t^{(1)}=1$.

FISTA converges as long as the gradient descent step size $\mu$ satisfies
\begin{equation}
0< \mu < \frac{1}{ \norm{\mathbf{A^{\dagger}} \cdot \mathbf{\Sigma}^{-1} \cdot \mathbf{A} }},
\end{equation}
where the operation $\norm{\bullet}$ returns the spectrum norm of the input
matrix. The spectral norm is estimated by simulating a large number of random
vectors with $l^2$ norms equal one with different realizations. The matrix
operator $\mathbf{A^{\dagger}} \cdot \mathbf{\Sigma}^{-1} \cdot \mathbf{A}$ is
subsequently applied to each random vector to yield a corresponding transformed
vector. The spectral norm is determined as the maximum $l^2$ norm of the
transformed vectors.

As summarized in the below, we first initialize the projection coefficients as
zero, and use FISTA to find the global minimum of the LASSO loss function.
Thus-obtained global minimum is the preliminary one. We then use the
preliminary LASSO estimation to weight the coefficients and calculate the
adaptive LASSO loss function. Finally, we set the preliminary LASSO estimation
as the ``warm'' start of the adaptive LASSO estimation, and FISTA again to find
the global minimum of the adaptive LASSO loss function, which is our final
solution.

\begin{algorithm}[H]
\renewcommand{\thealgorithm}{}
\label{alg-1}
\caption{Our Algorithm}
\begin{algorithmic}[1]
\INPUT $\gamma$: Pixelized complex $3$D shear field
\OUTPUT  $\delta$: $3$D array of density contrast
\STATE Normalize column vectors of $\mathbf{A}$
\STATE Estimate step size $\mu$ and $\mathbf{\Sigma}$
\STATE \textbf{Initialization:}
\STATE $x'^{(1)} = 0$
\STATE $\hat{w}=1$
\STATE $t^{(1)}=1$,$i=1$, $j=1$
\WHILE{$j \leq 2$}
    \WHILE{$i \leq N_{\rm{iter}}$}
        \STATE \# soft thresholding
        \STATE $x'^{(n+1)}_{i}=\mathrm{ST}_{\hat{w}\lambda} \left(x'^{(n)}_{i} -\mu \partial_i G(x'^{(n)})\right)$
        \STATE \# FISTA algorithm
        \STATE $t^{(n+1)}=\frac{1+\sqrt{1+4(t^{(n)})^2}}{2}$
        \STATE $x'^{(n+1)} \leftarrow x'^{(x+1)}+ \frac{t^{(n)}-1}{t^{(n+1)}}(x'^{(n+1)}-x'^{(n)})$
        \STATE $i=i+1$
    \ENDWHILE
\STATE \textbf{Re-initialization:}
\STATE $\hat{w}=\abs{\hat{x'}_{\rm{LASSO}}}^{-2}$, $\lambda \leftarrow
\lambda_{\rm{ada}}$
\STATE $\hat{x'}^{(1)} = x'^{(N_{\rm{iter}})}$
\STATE $t^{(1)}=1$, $i=1$
\STATE $j=j+1$
\ENDWHILE
\STATE $\delta=\mathbf{\Phi}\mathcal{N}^{-\frac{1}{2}}x'^{(N_{\rm{iter}})}$
\end{algorithmic}
\end{algorithm}

\section{Cluster Detection}
\label{sec_Test}

We simulate weak-lensing shear induced by NFW halos with various masses and
redshifts. The generated shear fields are used to distort the HSC mock galaxy
shapes with different realizations of shear measurement and photo-$z$
uncertainties (Section~\ref{subsec_Sims}).

We then test our algorithm using the mock catalogs with varying the penalty
parameter in our model (Section~\ref{subsec_test-nfw}). We also compare our
method that uses the NFW dictionary with one using the point mass dictionary
(Section~\ref{subsec_test-pm}).

\subsection{Simulations}
\label{subsec_Sims}
We use halos with a variety of masses and redshifts in the range $10^{14}
\text{h}^{-1}M_{\odot} < M < 10^{15} \text{h}^{-1}M_{\odot}$, and $0.05 < z <
0.85$, respectively. We divide the parameter space into eight redshift bins and
eight mass bins with equal separation. We randomly shift the input halo
redshift and halo mass from the bin center by a small amount in order to avoid
repeatedly sampling at the exact same halo mass and redshift.

The concentration parameter $c_{\rm h}$ of a NFW halo is determined as a
function of the halo mass ($M_{200}$) and redshift ($z_{\rm h}$) according to
\citet{c-M_Magneticum-Ragagnin2019}:
\begin{equation}
c_{\rm h}=6.02\times\left(\frac{M_{200}}{10^{13} M_{\odot}}\right)^{-0.12}
    \left(\frac{1.47}{1.+z_{\rm h}}\right)^{0.16}\,.
\end{equation}
The weak-lensing shear fields of the NFW halos are calculated according to
\citet{haloModel-TJ2003-3pt}. The shear distortions are applied to one hundred
realizations of galaxy catalogs with the HSC-like shear measurement error and
photo-$z$ uncertainty.

\begin{figure}
\centering
\includegraphics[width=0.45\textwidth]{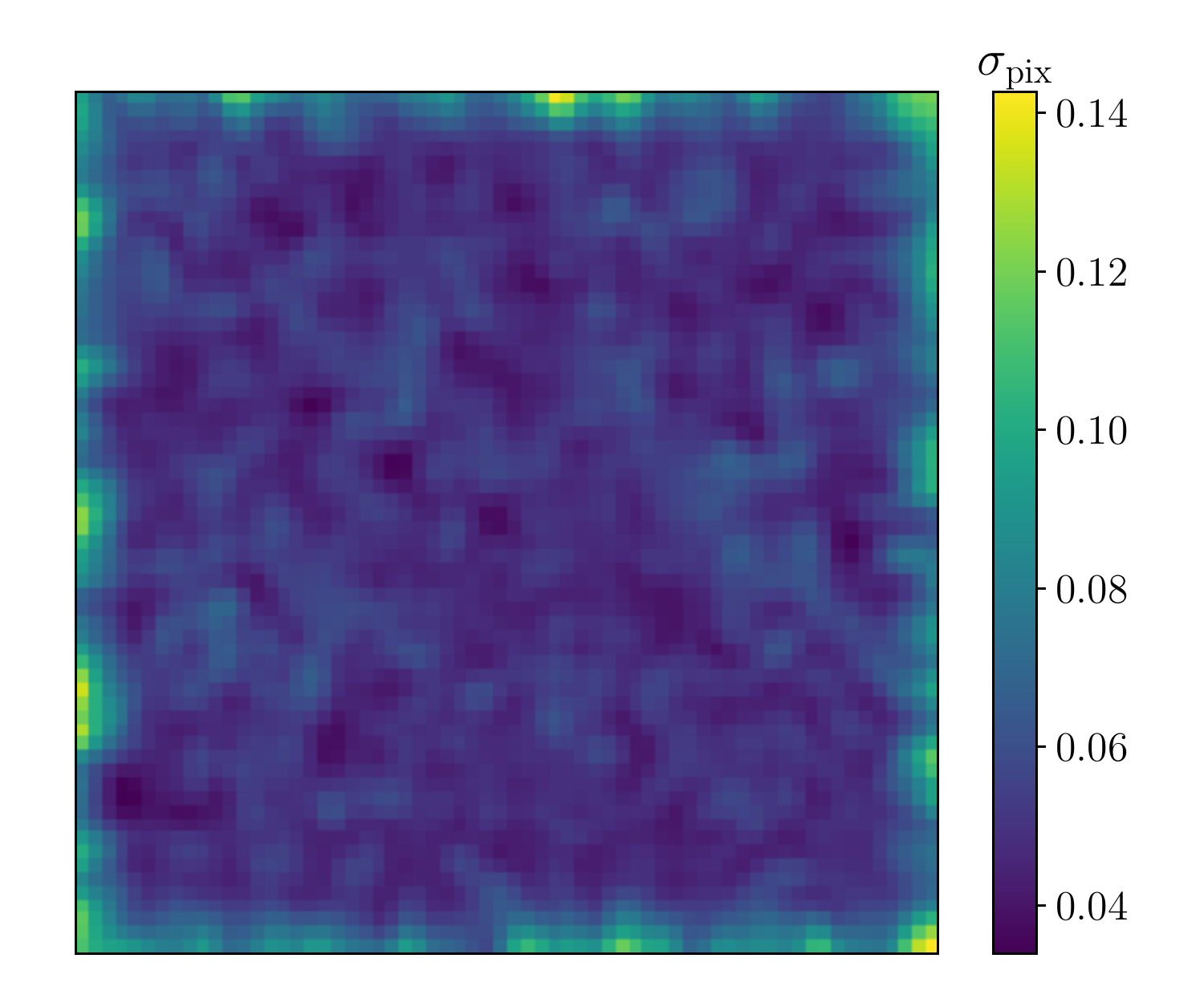}
\caption{
    The standard deviation pixel map of the HSC-like shear measurement error
    for the fifth source galaxy bin ($0.69 \leq z < 0.80$).
    }
    \label{fig_noistdmap}
\end{figure}

The mock galaxy catalogs are generated using the HSC S16A shear catalog
\citep{HSC1-catalog}. We use the galaxies in a 1 square degree field at the
center of tract 9347 \citep{HSC1-data}. The average galaxy number density in
this region is $22.94~\rm{arcmin}^{-2}$. The positions of galaxies are
randomized to distribute homogeneously in the one-square degree stamp. We
randomly assign redshift for each galaxy following the $\MLZ$ photo-$z$
probability distribution function \citep{HSC1-photoz}.

We simulate the HSC-like shear measurement error due to shape noise and sky
variance with different realizations by randomly rotating the galaxies in the
shear catalog. The shear measurement error on individual galaxy level and
individual pixel level are demonstrated in Figure~\ref{fig_noiseHistogram}. The
standard deviation map of the noise is demonstrated in
Figure~\ref{fig_noistdmap}.

\subsection{NFW atoms}
\label{subsec_test-nfw}
\begin{figure*}[!t]
\begin{minipage}[c]{1.0\columnwidth}
\begin{center}
    \includegraphics[width=1.0\textwidth]{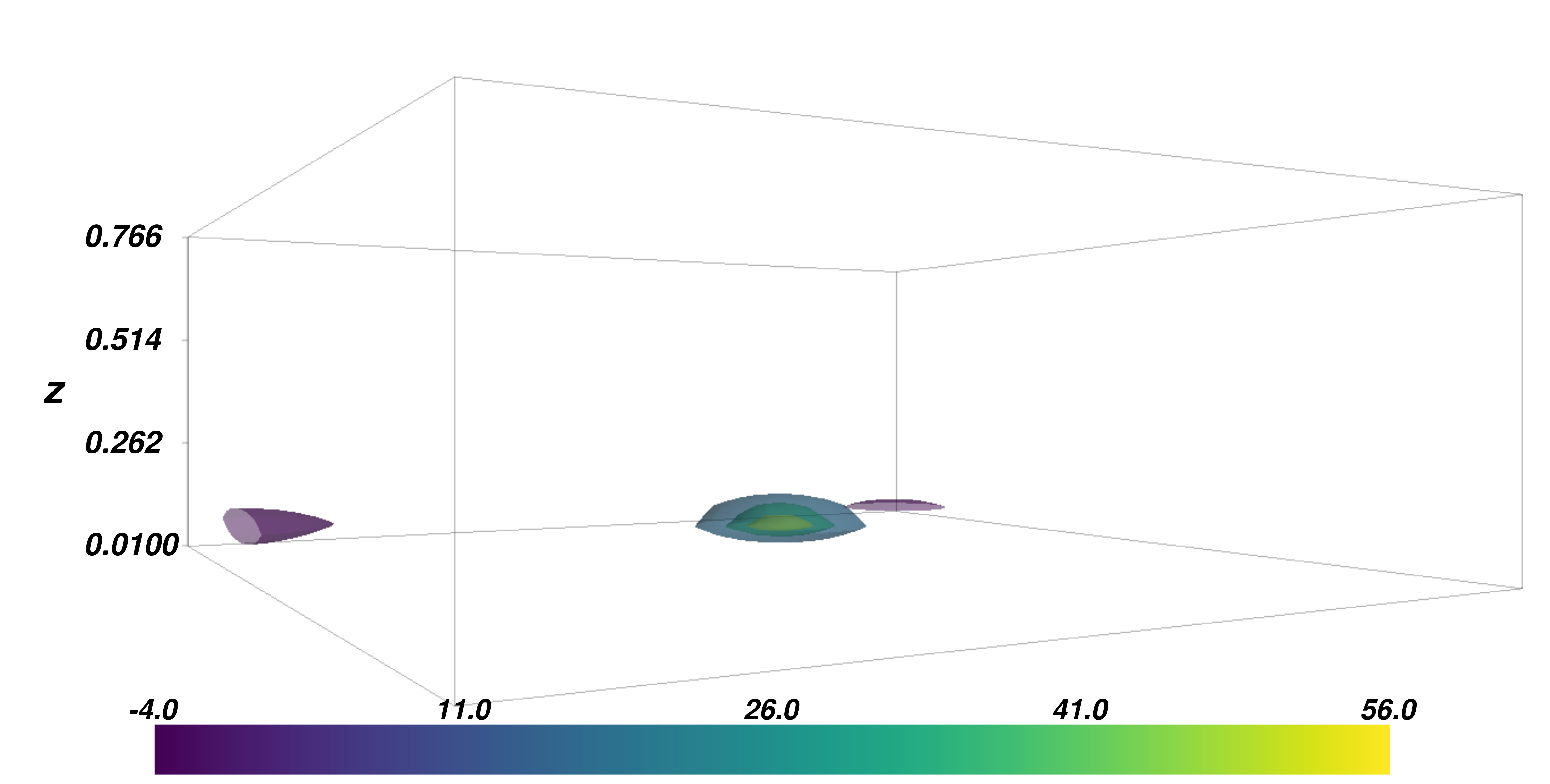}
    \includegraphics[width=0.9\textwidth]{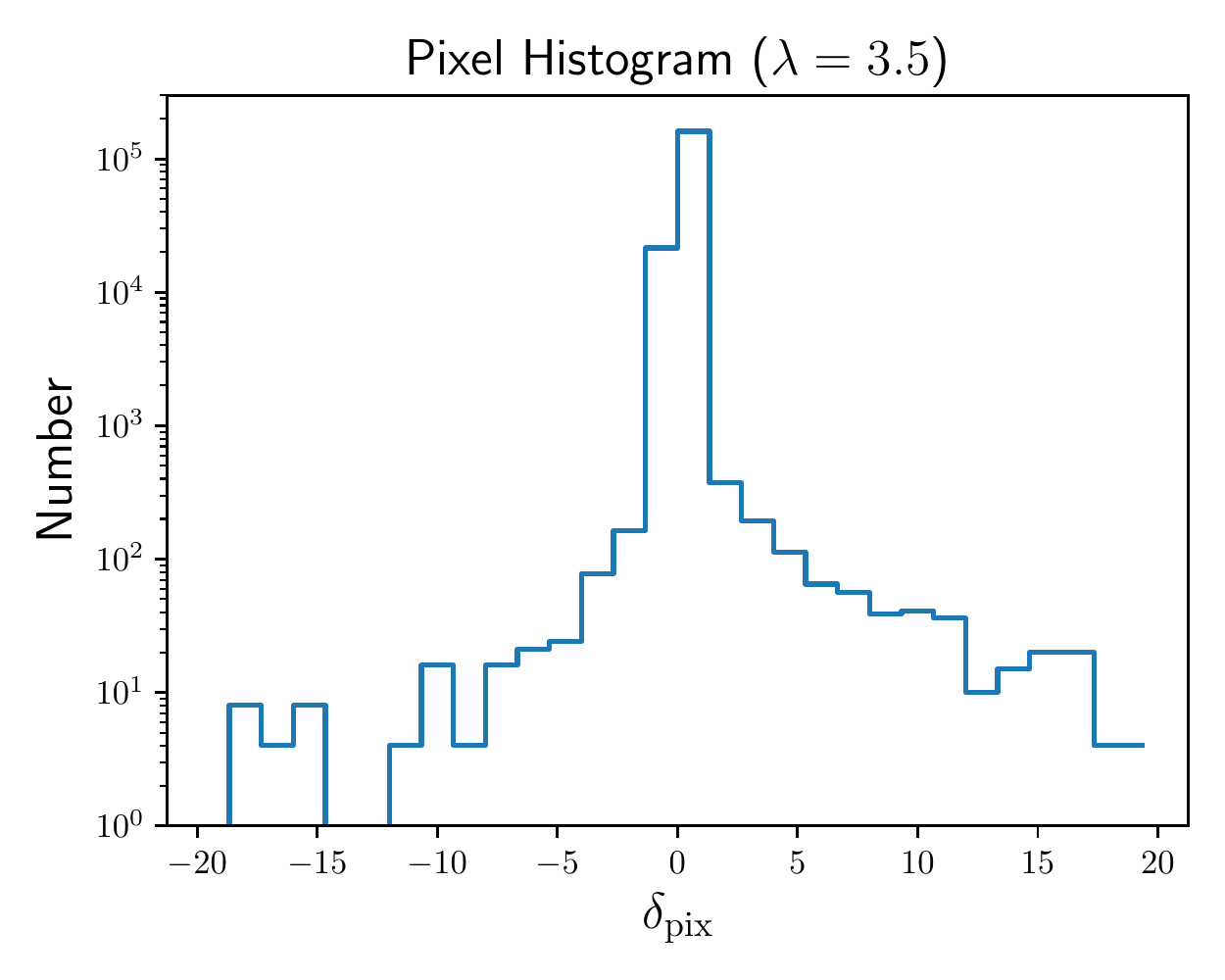}
\end{center}
\end{minipage}
\begin{minipage}[c]{1.0\columnwidth}
\begin{center}
    \includegraphics[width=1.0\textwidth]{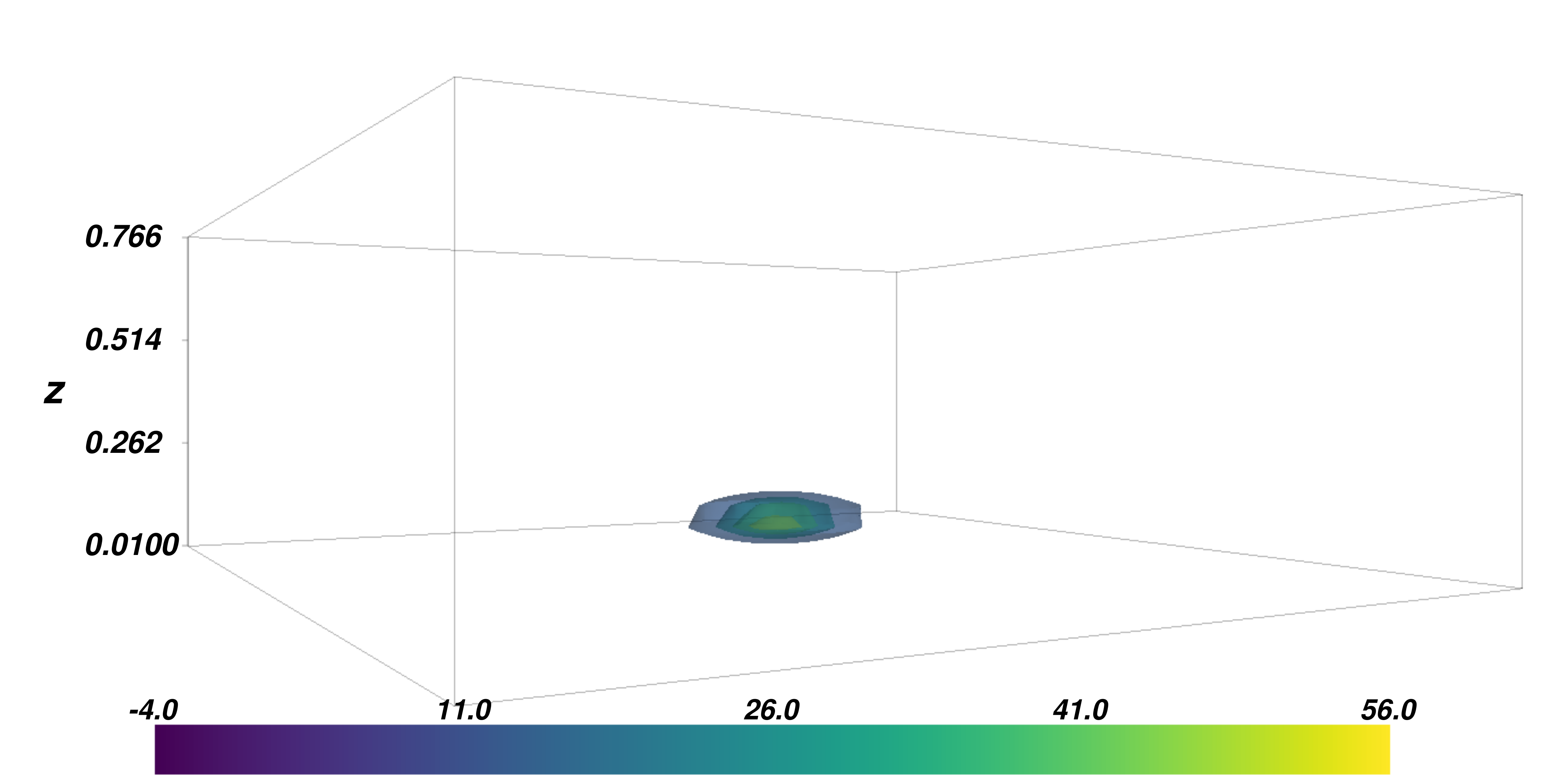}
    \includegraphics[width=0.9\textwidth]{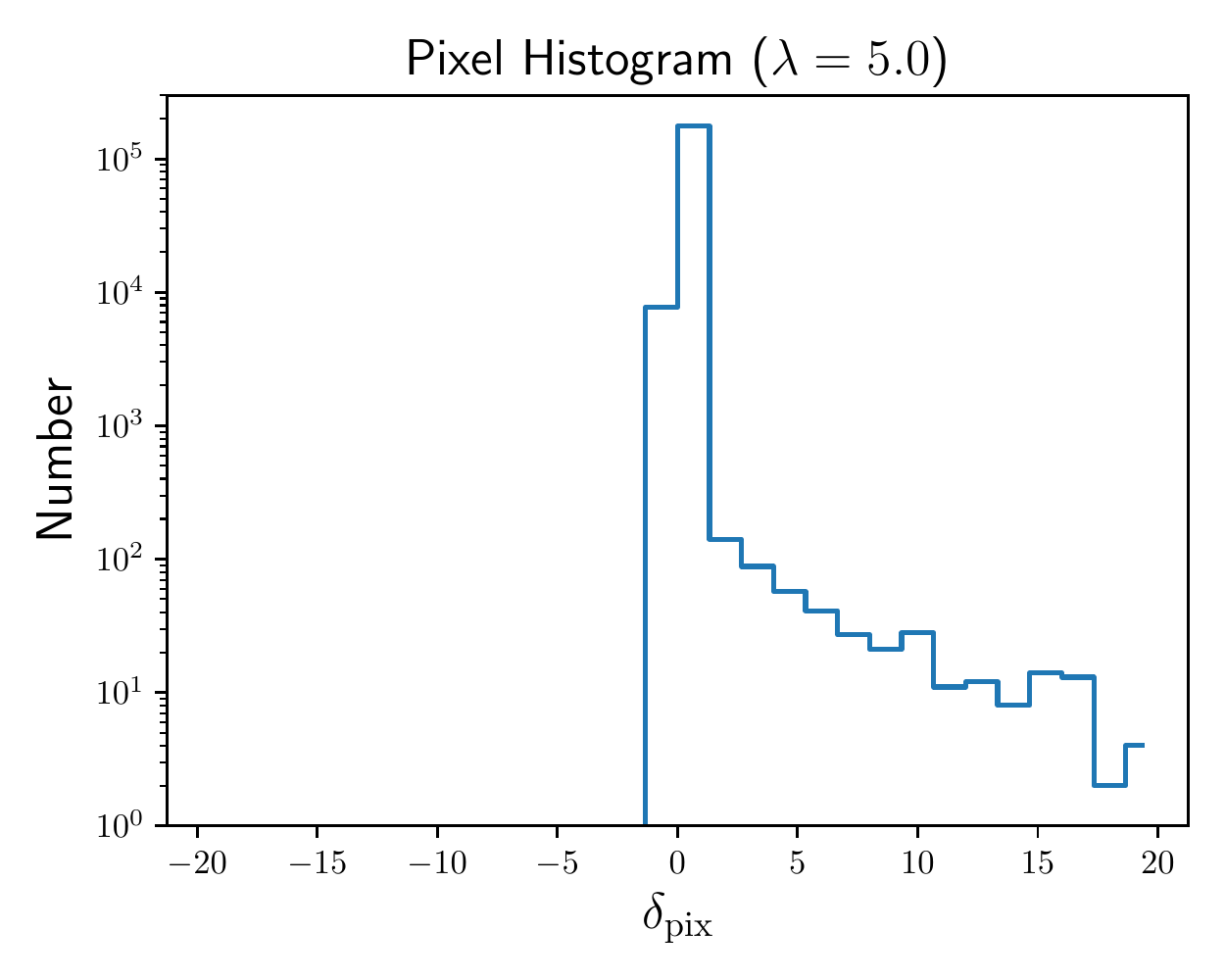}
\end{center}
\end{minipage}
\caption{
    The upper panels show the density maps reconstructed from a mock galaxy
    shear catalog, which includes shear measurement and photo-$z$
    uncertainties, using our algorithm with penalty parameters $\lambda=3.5$
    (left) and $\lambda=5.0$ (right). The axes and color maps have the same
    meanings as Figure~\ref{fig_LASSOVsadaLasso}. The lower panels show the
    corresponding number histograms of pixel values. The input halo mass is
    $M_{200}=10^{15.02} ~\text{h}^{-1}M_{\odot}$, and its redshift is $z=0.164\,$.
    }
    \label{fig_NFW3D}
\end{figure*}

In this section, we test the performance of our algorithm by adopting models
where the matter density field is represented by multiscale NFW atoms. The
dictionary is constructed with three frames of different NFW scale radii in
comoving coordinate: $0.12~\text{h}^{-1}$ Mpc, $0.24~\text{h}^{-1}$ Mpc, and
$0.36~\text{h}^{-1}$ Mpc. The truncation radii are set to four times the scale
radii for the atoms in the dictionary, i.e., we assume $c_{\rm h} =4$. Note
that each frame of our dictionary fixes the scale radius in {\it comoving}
coordinates and thus the NFW atoms have different angular sizes when placed at
different redshift.

We test the algorithm with varying the penalty parameter for the LASSO
estimation with $\lambda=3.5$, $4.0$, and $5.0$. The corresponding penalty
parameters for the final adaptive LASSO estimations are set to
$\lambda_{\rm{ad}}=\lambda^{\tau+1}$. Here, both the LASSO estimation and our
adaptive LASSO estimation select the pixels with signal-to-noise ratios greater
than $\lambda$ in each gradient descent iteration, and the local density is
estimated for the selected pixels with a shrinkage of the estimation amplitude.
The LASSO estimation shrinks the density amplitudes by $\lambda$ for every
selected pixels, whereas the adaptive LASSO estimation suppresses the shrinkage
if the preliminary estimation for the pixel is greater than $\lambda$, by
down-weighting their penalties, and otherwise it enhances the shrinkage.

We'd like to reconstruct with the resolution limit set by the Gaussian
smoothing kernel with a standard deviation of $1\farcm5$ and the $1\arcmin$
pixel scale as described in Sections~\ref{subsec_method_smoothing}. In
addition, we smooth the reconstructed density field with the same Gaussian
kernel in each lens redshift plane.

Figure~\ref{fig_NFW3D} shows the $3$D density maps reconstructed with different
penalty parameters for a halo with $M_{200}=10^{15.02} ~\text{h}^{-1}M_{\odot}$ at
redshift $0.164$. \xlrv{The corresponding pixel value histograms are shown in
Figure~\ref{fig_NFW3D}. Our adaptive LASSO algorithm assigns zero to a
fraction of the reconstructed pixels that are mostly noises, while
retaining strong signals.  The right panels demonstrate a case where almost
all of the pixels with pure noise are set to zero when a large penalty
parameteris adopted.} It is important to note that the reconstructed density
maps are not compromised by line of sight smearing, i.e. the reconstructed lens
is localized in redshift.

\begin{figure*}
\centering
\includegraphics[width=1.0\textwidth]{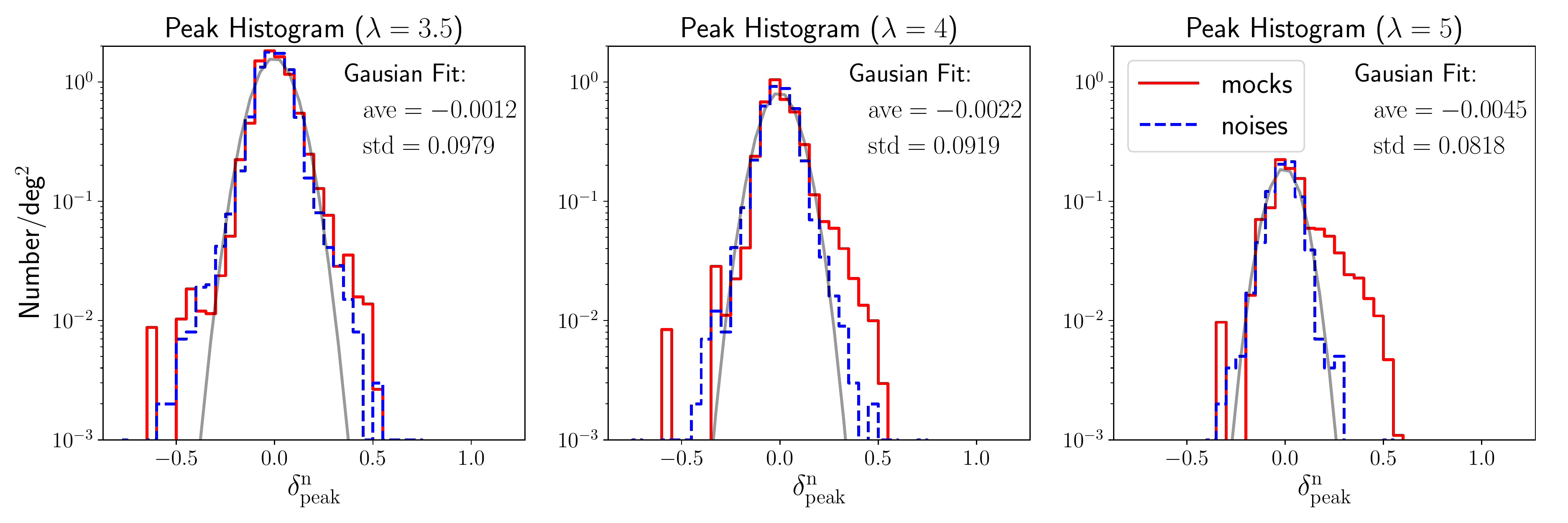}
\caption{\xlrv{
    The peak counts as a function of peak overdensity for maps reconstructed
    with different setups. We plot the peak number density per square-degree.
    }
    The solid red histograms show the results of the reconstructions from the
    mock shear catalogs with penalty parameters: $\lambda=3.5,4.0,5.0$, from
    left to right. The dashed blue histograms are the corresponding results of
    the reconstruction from $1000$ realizations of pure noise catalogs. The
    gray lines are the best-fit Gaussian distributions to the noise peak
    histograms.
    }
    \label{fig_peakHist}
\end{figure*}

Following \citet{LSS-massMap-Glimpse3D-Leonard2014}, we normalize the detected
peaks in the $l$th ($l=1...20$) lens redshift plane to account for the peak
amplitude difference arising from the difference in the norm of the lensing
kernels in different redshift bins:
\begin{equation}
    \delta^{\rm{n}}_{\rm{peak}}(\vec{\theta},z_l)=
    \delta_{\rm{peak}}(\vec{\theta},z_l)/\mathcal{R}_{l}^{\frac{1}{2}},
\end{equation}
where the normalization matrix is defined as
\begin{equation}
\mathcal{R}_{l}=\sum_s K^2(z_l,z_s)\,,
\end{equation}
where $K(z_l,z_s)$ is the lensing kernel. In Figure~\ref{fig_peakHist}, we show
the histograms of the normalized peaks with different penalty parameters.
There, we stack the histograms from $100$ realizations of all the halos sampled
in the redshift-mass plane. In addition, we generate $1000$ realizations of
pure noise catalogs and perform the reconstruction using the noise catalogs in
order to examine the noise properties. The histograms of the normalized peaks
detected from the pure noise catalogs are shown in Figure~\ref{fig_peakHist}
along with the best-fit Gaussian functions of the noise peak histograms.

We find that the number counts including both true and false peaks decrease as
the penalty parameter $\lambda$ increases. In addition, the standard deviation
of noise peaks decreases as $\lambda$ increases. As a result, with
$\lambda=5.0$, we find a clearer excess in the positive peak counts compared
with the noise peak histograms, especially at the high density contrast. This
is expected because a higher penalty parameter prefers a sparser solution; more
peaks originating from the noise are removed than those from real clusters at
the high density contrast.

\begin{figure*}
\centering
\includegraphics[width=1.0\textwidth]{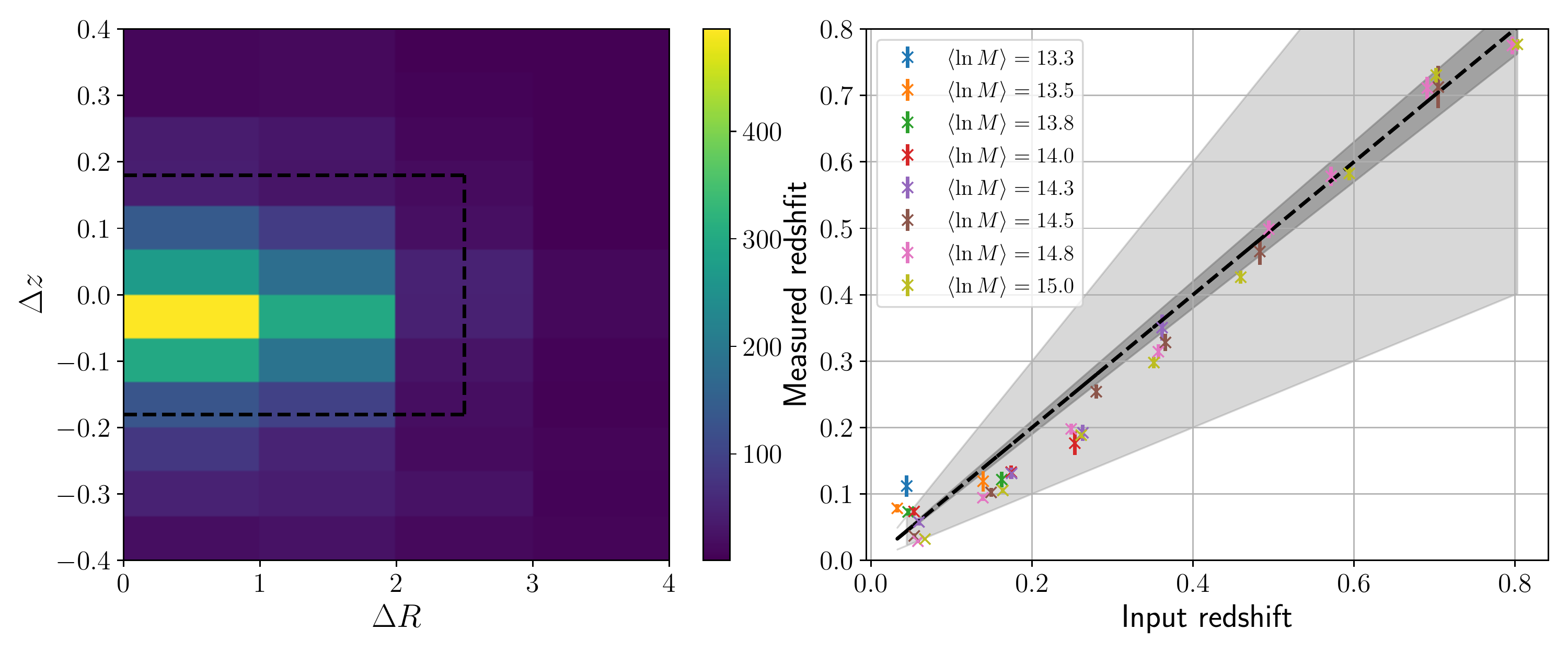}
\caption{
    The left panel shows the stacked $2$D distribution of the deviations of
    detected peak positions from the centers of the corresponding input halos.
    The $x$-axis is for the deviated distance in the transverse plane, and the
    $y$-axis is for the deviation in redshift. In each simulation, the positive
    peak inside the dashed black box with the minimal offset (in the pixel
    unit) from the position of the input halo is taken as ``true'' detection.
    The right panel shows the redshift deviation of detected peaks. The
    $x$-axis is the input halo redshifts, and the $y$-axis is the redshift of
    the detected peak. The cross-points denote the average of the detected
    peaks for each halo over different noise realizations, and the error-bars
    indicate the uncertainties of the averages. The deep gray area indicates
    relative redshift bias less than $0.05$, and the light gray area for
    relative redshift bias less than $0.5$. These results are based on our
    reconstruction with the NFW dictionary with $\lambda=3.5\,$.
    }
    \label{fig_detoffsets}
\end{figure*}

We demonstrate the position offsets of the detected halos compared to the input
positions in Figure~\ref{fig_detoffsets}. The left panel shows the $2$D number
histograms stacked over all the simulations as a function of the angular
distance and the redshift offsets. We see clear clustering of the peaks close
to the position of the input halo on the stacked number histogram. For each
simulation, we identify positive peaks closest to the input position (in the
pixel unit). If a closest peak is located inside the region denoted with the
dashed box in the left panel of Figure~\ref{fig_detoffsets}, we regard it as a
``true'' peak detection. Other identified peaks, which include both positive
and negative peaks, are judged to be false detection.

The right panel of Figure~\ref{fig_detoffsets} shows the estimated redshift of
the ``true'' detections averaged over the simulations (with different noise
realizations) for each halo as a function of the input redshift. As shown, the
estimated redshifts are slightly lower than the ground truth by $\Delta z \sim
0.03$ for halos in the low-redshift range ($z\leq 0.4$). For halos at
$0.4<z\leq 0.85$, the relative redshift bias is below $0.5\%$. The standard
deviation of the redshift estimation is $0.092$.

\begin{figure*}
\begin{center}
    \includegraphics[width=1.0\textwidth]{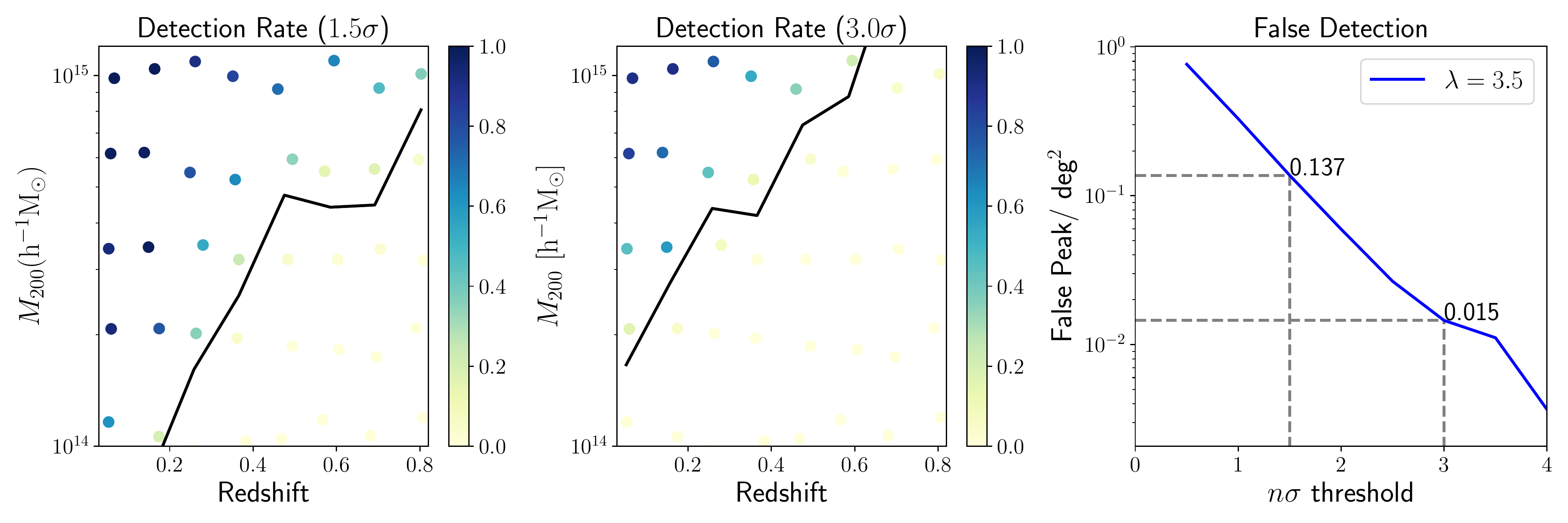}
\end{center}
\caption{
    The detection rates and false peak densities for different detection
    thresholds. The left (middle) panel shows the halo detection rates for
    detection threshold that equals $1.5\sigma$ ($3.0\sigma$). The black lines
    indicate the contours for detection rate $0.1\,$. The right panel shows the
    density of false peaks as a function of detection threshold. The penalty
    parameter is set to $\lambda=3.5\,$.
    }
    \label{fig_finalres_lbd35}
\end{figure*}

\begin{figure*}
\begin{center}
    \includegraphics[width=1.0\textwidth]{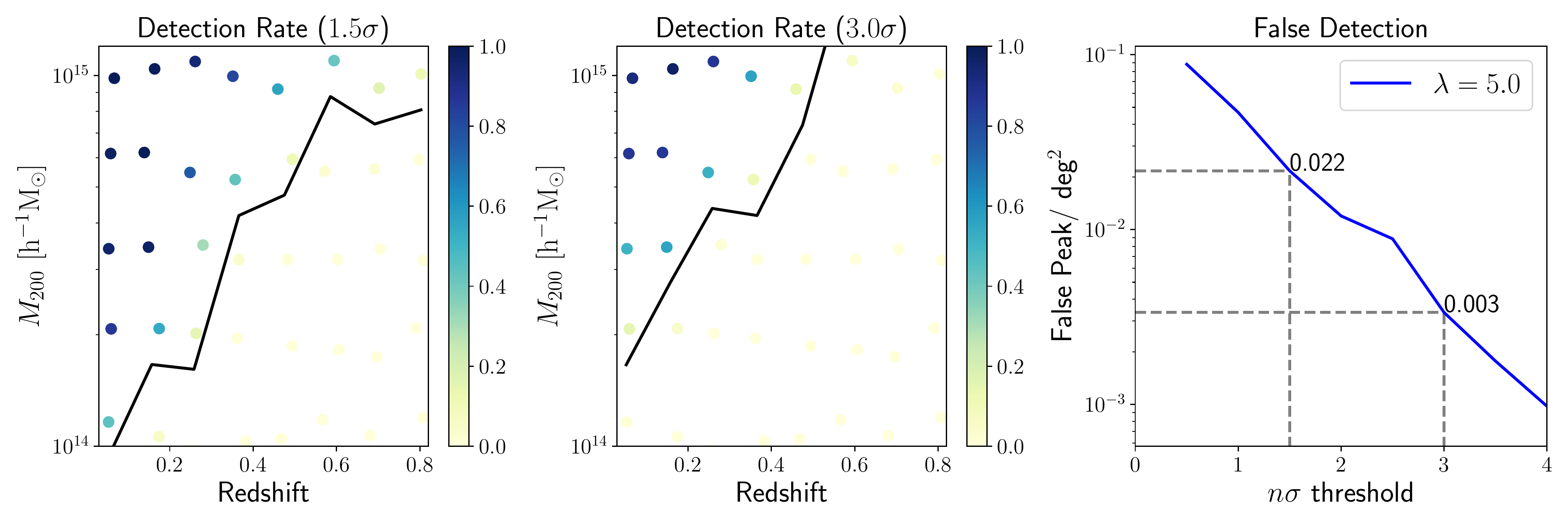}
\end{center}
\caption{
    As for Figure~\ref{fig_finalres_lbd35}, but for the penalty parameter
    $\lambda=5.0\,$.
    }
    \label{fig_finalres_lbd50}
\end{figure*}

In order to reduce the number of false detections, we select peaks as galaxy
clusters if the peak values are greater than an empirically determined
threshold. The threshold is set in units of the standard deviation of the noise
peaks. We use different detection thresholds ($1.5\sigma$ and $3.0\sigma$) to
detect galaxy clusters from the mass maps reconstructed with different penalty
parameters. Figures~\ref{fig_finalres_lbd35} and \ref{fig_finalres_lbd50} show
the detection rates for halos in the redshift-mass plane for different penalty
parameters ($\lambda=3.5$ and $\lambda=5.0$). The corresponding number of false
detections per square degree as a function of detection threshold are also
demonstrated in the figures.

As shown, the false peak density is successfully reduced for relatively large
detection threshold, but the detection rate of halo also decreases. After a few
experiments, we have decided to set the detection threshold to $1.5\sigma$ and
set the penalty parameter $\lambda$ to $5.0$ since the combination suppresses
the false detection to $0.022$ while keeping a high halo detection rate. In
summary, our method is able to detect halos with minimal mass of $10^{14.0}
\text{h}^{-1}M_{\odot}$, $10^{14.7} \text{h}^{-1}M_{\odot}$, $10^{15.0}
\text{h}^{-1}M_{\odot}$ for the low ($z<0.3$), median ($0.3\leq z< 0.6$) and
high ($0.6\leq z< 0.85$) redshift ranges, respectively.

\begin{figure}
\begin{center}
\includegraphics[width=0.45\textwidth]{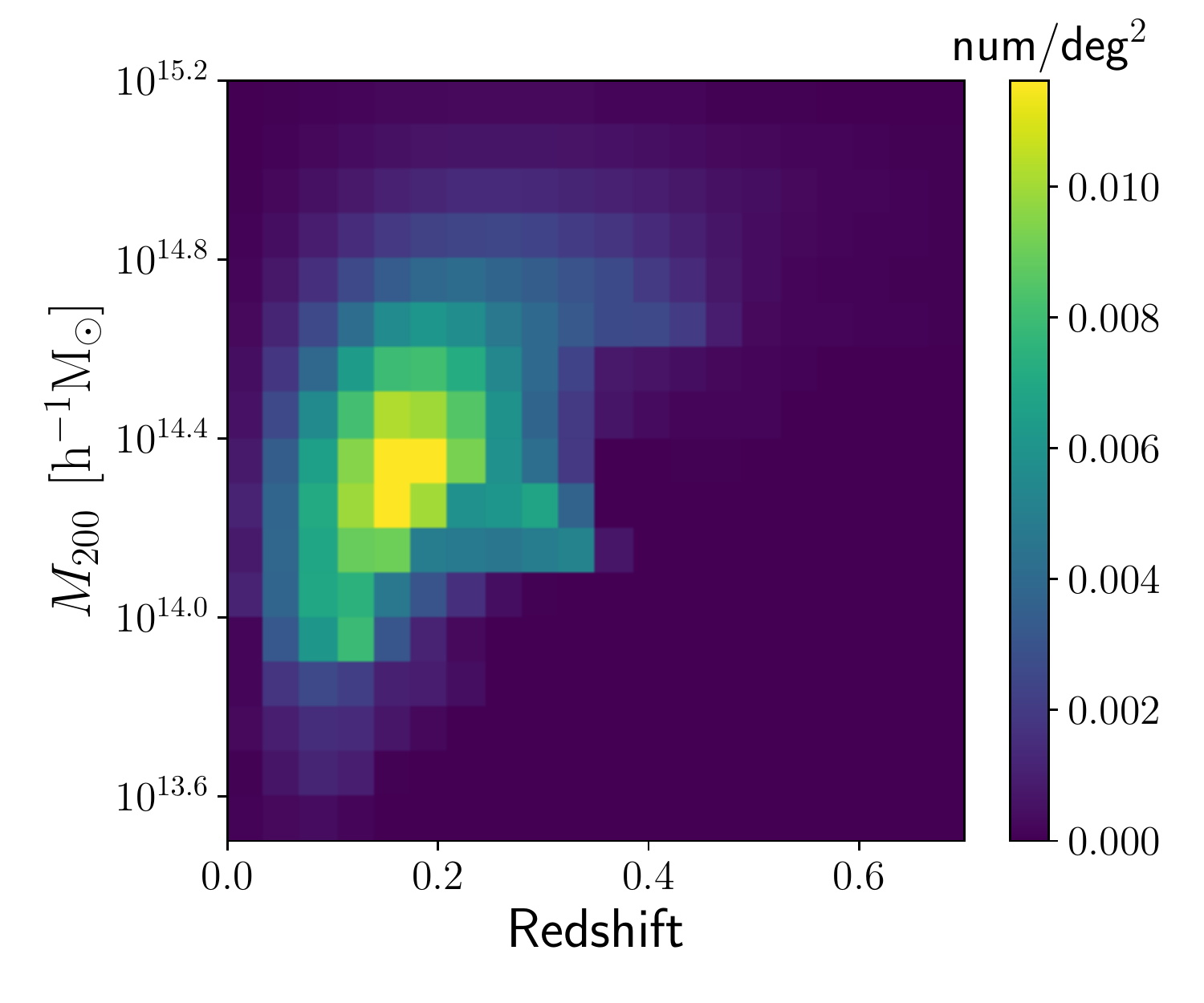}
\end{center}
\caption{
    The expected number density of detected clusters per square degree as a
    function of halo redshift ($x$-axis) and halo mass ($y$-axis). The number
    density in total is $0.49~\deg^{-2}\,$.
    }
    \label{fig_detNum}
\end{figure}

Using the detection rate measured from our simulations, we are able to predict
the number density of detected clusters by assuming the halo mass function of
\citet{haloMass-Tinker2008}. We use HMF \citep{hmf-Murray2013}, an open-source
package\footnote{\url{https://github.com/halomod/hmf}}, to calculate the halo
mass function. The predicted halo detection number density for the setup
$\lambda=5$ and $1.5\sigma$ detection threshold is shown in
Figure~\ref{fig_detNum}. The resulting cluster number density is $0.49$
$\deg^{-2}$, which is much higher than the false detection rate of 0.022
deg$^{-2}$. This cluster number density corresponds to $78.4$ clusters for the
first-year HSC shear catalog \citep{HSC1-catalog} with a survey area of $\sim
160 \deg^2$. The expected number of detection is slightly higher than the
number of $2$D cluster detections ($63$ detected clusters) for the first year
HSC shear catalog \citep{HSC1-massMap-cluster}. Furthermore, our $3$D detection
method provides an accurate redshift estimation for individual clusters. In
contrast, the redshift information is not provided from the $2$D mass map
reconstruction.

\subsection{Point mass atoms}
\label{subsec_test-pm}

We perform an additional test by substituting the default NFW dictionary with
point-mass. This test may indicate some certain limitation of our method when
applied to a case with very compact (point) objects, which however is unlikely
to happen in actual lensing observations.

We set the penalty parameter for the preliminary LASSO to $\lambda = 3.5$ and
$5.0$. \citet{structureAdaLasso-Pramanik2020} propose to incorporate group
information into different adaptive LASSO penalty weights by setting the
weights for projection coefficients in a group to the average of the adaptive
weights in the same group. We assume that the neighboring pixels in the same
redshift plane belong to the same structural group (e.g., galaxy cluster and
void), and smooth the amplitude of the preliminary LASSO estimation in each
lens redshift plane with a top-hat filter of comoving diameter $r_{\rm c} =
0.25~\text{h}^{-1}$ Mpc. Let us denote the amplitude of the preliminary LASSO
estimation with the smoothing as $\abs{\hat{x}^{\rm{LASSO}}}_{\rm{sm}}$, and
adopt the penalty weights given by
$\hat{w}=1/\abs{\hat{x}^{\rm{LASSO}}}_{\rm{sm}}^{\tau}$.

\begin{figure*}
\centering
\begin{minipage}[c]{1.0\columnwidth}
\begin{center}
    \includegraphics[width=1.0\textwidth]{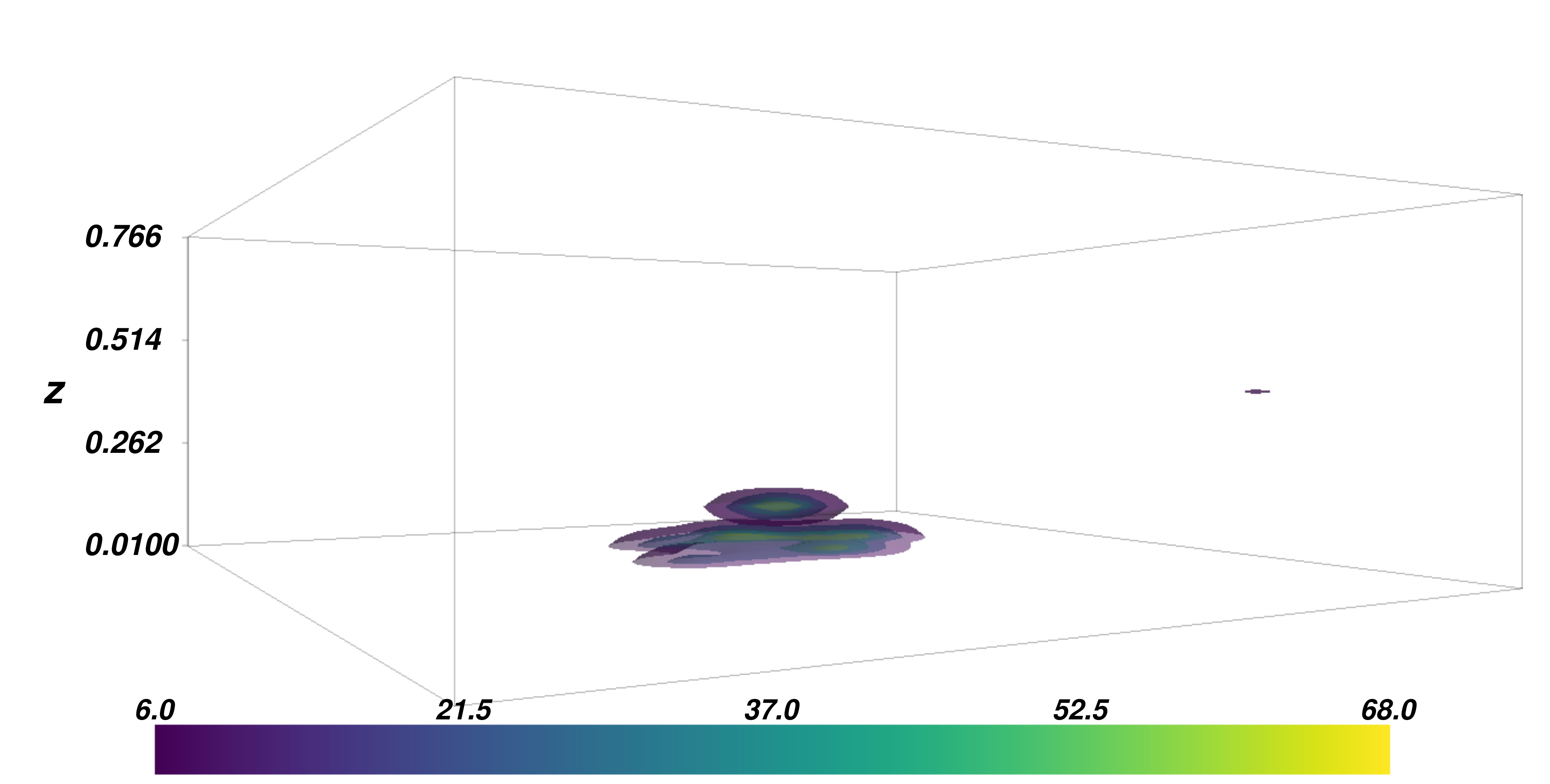}
\end{center}
\end{minipage}
\begin{minipage}[c]{1.0\columnwidth}
\begin{center}
    \includegraphics[width=1.0\textwidth]{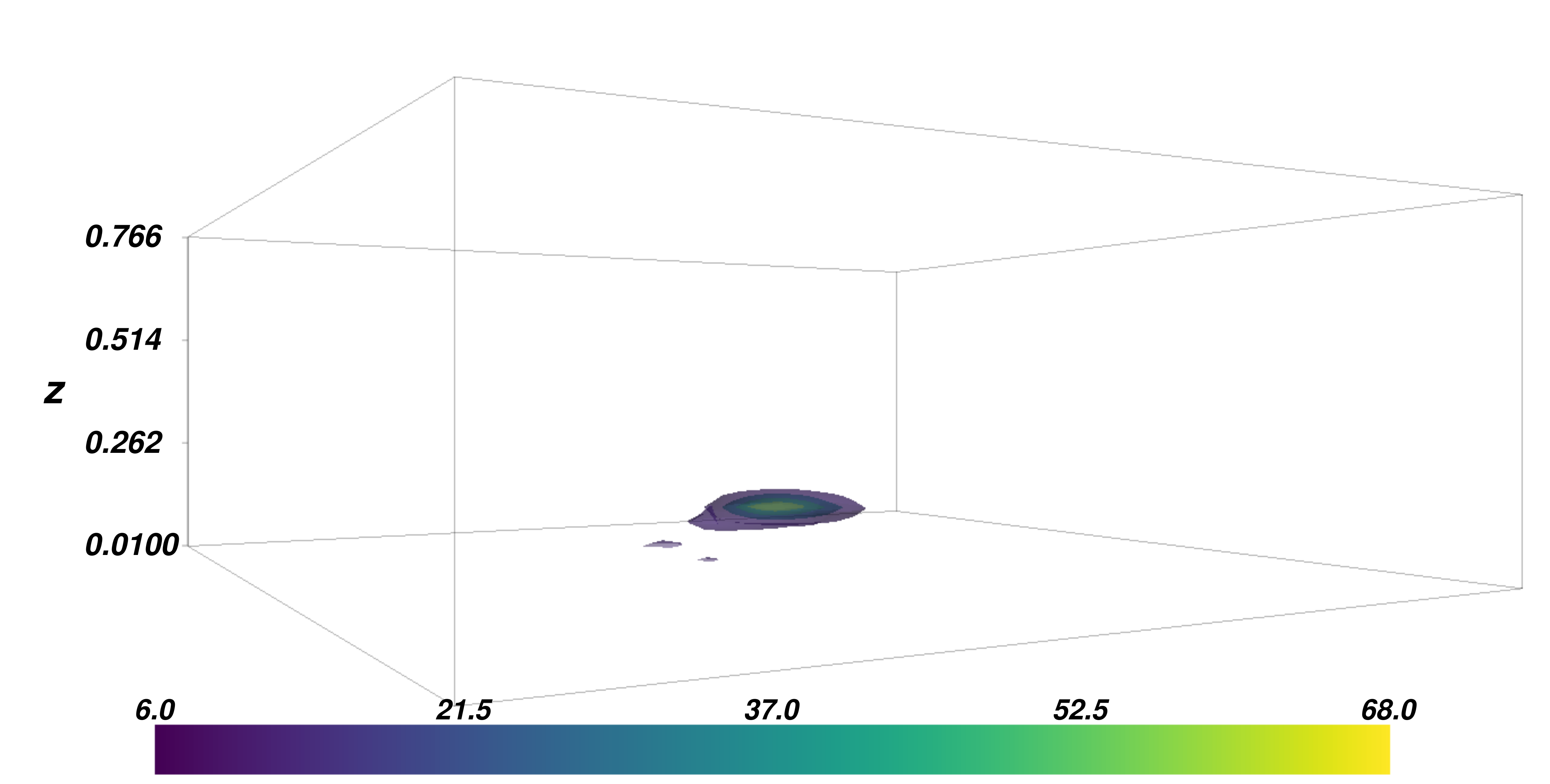}
\end{center}
\end{minipage}
\caption{
    The density maps reconstructed from the mock galaxy shape catalog with the
    point-mass dictionary. The penalty parameter is $\lambda=3.5$ (left) and
    $\lambda=5.0$ (right). The axes and color maps have the same meanings as
    Figure~\ref{fig_LASSOVsadaLasso}. The input halo mass is
    $M_{200}=10^{15.02} ~\text{h}^{-1}M_{\odot}$, and its redshift is $z=0.164$.
    }
    \label{fig_PM3D}
\end{figure*}

Figure~\ref{fig_PM3D} shows the reconstruction result with the point-mass
dictionary. Interestingly, several ``discrete'' masses are assigned to at
different redshift bins in the neighboring region of the true halo center. In
contrast, as we have seen in Figure~\ref{fig_NFW3D}, the NFW dictionary manages
to recover a consistent mass distribution. The problem of the point-mass
dictionary originates from the fact that the profile of the point-mass atom in
the transverse plane is much more compact than the profile of the input halo,
especially when placed at low redshifts.

\section{Summary}
\label{sec_Sum}

We have developed a novel method to generate high-resolution $3$D density maps
from weak-lensing shear measurement with photometric redshift information. A
key improvement over previous similar methods is that we represent a $3$D
density field by a collection of NFW atoms with different physical sizes. With
a prior assumption that the clumpy mass distribution is sparse in $3$D, we
reconstruct the density map using the adaptive LASSO algorithm
\citep{AdaLASSO-Zou2006}. We show that adopting the standard LASSO algorithm
results in significant smearing of structure in the line of sight direction
even in the absence of galaxy shape noise and photometric redshift
uncertainties. Our adaptive LASSO algorithm efficiently reduces the smearing of
structure.

We have examined the performance of cluster detection with the reconstructed
$3$D mass maps using mock catalogs that apply shear distortions from isolated
halos to galaxies with HSC-like shapes and photo-$z$ uncertainties.  Under the
realistic conditions, our method is able to detect halo with minimal mass
limits of $10^{14.0} \text{h}^{-1}M_{\odot}$, $10^{14.7}
\text{h}^{-1}M_{\odot}$, $10^{15.0} \text{h}^{-1}M_{\odot}$ at low ($z<0.3$),
median ($0.3\leq z< 0.6$) and high ($0.6\leq z< 0.85$) redshifts, respectively,
with an average false detection of 0.022 deg$^{-2}$. The estimated redshifts of
the clusters detected in the reconstructed mass maps are slightly lower than
the true redshift by $\Delta z \sim 0.03$ for halos at low redshifts ($z\leq
0.4$). The relative redshift bias is below $0.5\%$ for halos at $0.4<z\leq
0.85$, and the standard deviation of the redshift estimation is $0.092$.

We will apply the newly developed $3$D mass map reconstruction technique to the
wide-field data from HSC survey \citep[e.g.,][]{HSC1-catalog,FPFSHSC1-Li2020}
and perform galaxy cluster detection in our future work.

\section*{Acknowledgements}
We thank Yin Li and Jiaxin Han for useful discussions. X.L. was supported by
Global Science Graduate Course (GSGC) program of University of Tokyo and JSPS
KAKENHI (JP19J22222).

This work was supported in part by Japan Science and Technology Agency (JST)
CREST JPMHCR1414, and by JST AIP Acceleration Research grant No. JP20317829,
and by the World Premier International Research Center Initiative (WPI
Initiative), MEXT, Japan, and JSPS KAKENHI grant Nos. JP18K03693, JP20H00181,
JP20H05856.

\bibliographystyle{aasjournal}
\bibliography{citation}

\end{document}